\documentclass[runningheads]{llncs}
\usepackage{graphicx}
\usepackage{makeidx}         

\usepackage{xcolor}     
\usepackage{amsmath}    

\begin{document}
\title{Entropy and the City:\\Origins, trajectories and explorations of the concept in urban science
}
\titlerunning{Entropy and the City}
%
\author{Vinicius M. Netto\inst{1}
\and Otavio M. Peres\inst{2}
\and Caio Cacholas\inst{3}
}
\authorrunning{Netto et al.}
%
\institute{Research Centre for Territory, Transports and Environment (CITTA), Faculty of Engineering (FEUP), University of Porto, Portugal. \email{vmnetto@fe.up.pt} \and
Faculty of Architecture and Urbanism (FAUrb), Federal University of Pelotas (UFPel), Brazil.
\email{otavio.peres@ufpel.edu.br} \and Graduate Programme of Architectural and Urban Studies (PPGAU), Federal Fluminense University(UFF), Brazil. \email{caiocacholas@id.uff.br}}
\maketitle              
\begin{abstract}
Entropy is arguably one of the most powerful concepts to understand the world, from the behavior of molecules to the expansion of the universe, from how life emerges to how hybrid complex systems like cities come into being and continue existing. Yet, despite its widespread application, it is also one of the most misunderstood concepts across the sciences. 
This chapter seeks to demystify entropy and its main interpretations, along with some of its explorations in the context of cities.
It first establishes the foundations of the concept by describing its trajectory since its inception in thermodynamics and statistical mechanics in the 19th century, its different incarnations from Boltzmann's pioneering formulation and Shannon’s information theory to its absorption in biology and the social sciences, until it reaches a nascent urban science in the 1960s. 
The chapter then identifies some of the main domains in which entropy has been explored to understand cities as complex systems -- from entropy-maximizing models of spatial interaction and applications as a measure of urban form, diversity, and complexity to a tool for understanding conditions of self-organization and urban sustainability.
\keywords{cities \and entropy  \and urban science}
\end{abstract}
%


\section{Introduction} \label{introduction_1}
\begin{flushright} \emph{Now, entropy is a very strange concept.} \\Ilya Prigogine~\cite{prigogine1989entropy} \end{flushright}

Entropy is seen everywhere -- from the behavior of molecules to the evolution of the universe, from how life emerges to how hybrid complex systems like cities~\cite{portugali2021cities} come into being and continue to exist. Entropy is arguably one of the most powerful concepts to understand the world. Indeed, there are few concepts that have more widespread applicability in physical sciences than entropy and its irresistible quality in defining the structure and behavior of systems in different scientific fields~\cite{battySpatialEntropy1974a}. Yet, it is also one of the most misunderstood concepts in science and common sense alike. It is so for several reasons, both in general and within the specific context of urban studies and the science of cities. It is a complex property rooted in thermodynamics, statistical mechanics, and information theory, which can be highly abstract and counter-intuitive, involving mathematical formulations and probabilistic reasoning. 
The fact that the concept often involves leaps from micro to macro scales poses challenges to understanding how the statistical behavior of individual parts contributes to the macroscopic behavior of a system, including cities, where the behavior of individual components contributes to emergent properties at the city level. 

The frequent understanding of entropy as disorder can be misleading. While entropy is associated with the statistical measure of disorder in a thermodynamic system, this property does not directly translate into disorder in the everyday sense. The tremendous success of the concept in finding applications across a wide range of disciplines, from physics to urban studies, can make it difficult to reconcile its different interpretations and appreciate its universality. Finally, applying entropy to urban studies involves dealing with the multifaceted dynamics of cities, which are influenced by social, economic, and environmental factors. The complexity of urban systems can make it challenging to isolate and quantify the specific contributions of entropy, leading to potential misinterpretations. Entropy is applied in various fields of urban science, such as modelling spatial interactions, measuring diversity, assessing sustainability, and understanding self-organization. The diversity of applications can also contribute to confusion, as we may encounter entropy in different forms, materialities, and contexts.

This chapter seeks to demystify entropy and its main interpretations, along with some of its main explorations in the context of cities. It is intended as a means of understanding a famously hard-to-define concept while making sense of its plethora of applications. It aims to do so by identifying the main problems to which the concept has been put to use in research on cities, thus aiming to offer some semantic organization to an increasingly diverse area. It is intended as an introductory but comprehensive guide for those working in both qualitative and quantitative research in urban studies and the so-called ``new science of cities''. We begin by briefly elucidating the historical evolution of the concept, tracing its path from Clausius'~\cite{clausiusUeberVerschiedeneFuer1865} thermodynamic understanding and Boltzmann's~\cite{boltzmann1970weitere} groundbreaking work on statistical mechanics in 19th-century to Claude Shannon's~\cite{shannon1948} information theory. 
This foundation serves as a launching pad for its subsequent trajectories in biology and social sciences, filtering down into specialized fields of an emerging urban science that have gained momentum since the late 1960s. 
We shall unfold the multiple domains where the concept has been explored as a powerful lens to decipher the intricacies of cities -- namely, in the following applications.
\begin{itemize}
    \item[$\bullet$] Spatial interaction models of urban systems grounded in entropy-maximizing principles;
    \item[$\bullet$] The complexity and information latent in cities, i.e.\ how urban systems, like other dynamical systems, tend toward states of higher disorder and randomness associated with complexity and the amount of information in the built environment;
    \item[$\bullet$] Entropy as an approach to urban form, a lens to dissect the complexity inherent in the physical structures of cities, such as building footprints and street networks, as potential information signatures of distinct spatial cultures;
    \item[$\bullet$] Shannon entropy as a measure of urban diversity -- a metric essential for comprehending the intricate fabric that defines cities;
    \item[$\bullet$] Entropy as a tool to measure spatial segregation in urban landscapes;
    \item[$\bullet$] Entropy as a means to understand urban densities and the forces that shape the growth and sprawl of cities over time;
    \item[$\bullet$] Entropy as a crucial property in the self-organization of cities and their role in social dynamics and cooperation, the intricate processes that allow cities to evolve in response to internal and external stimuli and fluctuations; 
    \item[$\bullet$] Entropy as a means to assess urban sustainability -- the complex interplay between human activities, environmental factors, and the resilience of urban systems.
\end{itemize}

We build such a panorama of multifaceted properties, domains, and applications of entropy, aiming for both an intuitive and nuanced understanding of this concept and how it has been used to unveil the dynamics of cities and their ever-evolving social and material fabric. Like other chapters in this compendium, we shall explore the measures associated with such domains, along with some of the main empirical findings in the fields. We shall conclude the chapter with some critical and open questions in the field of entropy research in urban science and a brief summary of the main ideas explored.

\section{Historical trajectories of the concept} \label{historical_2}

\begin{flushright} \emph{The difficulties over the definition of entropy seem to originate from the fact that at one extreme, the concept is the basis of the second law of thermodynamics, which states that the entropy of a physical system must always increase, whereas, at the other extreme, the concept is used in defining the amount of information contained in a probability distribution. In fact, there is a somewhat tortuous route linking these two interpretations through arguments involving statistical physics. Yet the implications of the concept are still difficult to trace in any definitive sense} \\ Batty \cite{battySpatialEntropy1974a}
\end{flushright}

How did the concept and measure of entropy come to be related to the city? The study of entropy certainly cuts across many different research domains, from its original context in physics through the pioneering definitions of information theory and the life sciences to views of society, economies, and the evolution of cities. We bring here a brief account of such a rich trajectory of the concept, pointing out their connections with and explorations in urban studies and the science of cities.

\subsection{Origins in thermodynamics and statistical mechanics: Clausius and Boltzmann entropies} \label{hist_therm}

Clausius~\cite{clausiusUeberVerschiedeneFuer1865} introduced the concept of entropy in 1865 as a crucial idea in the Second Law of Thermodynamics. Etymologically, he derived the term in relation to the physical property of energy and the Greek word for change or transformation, ``trop\={e}''.
Clausius~\cite{clausiusUeberVerschiedeneFuer1865} defined entropy as the internal property that changes as heat energy moves around within a system -- ``the differential of a quantity which depends on the configuration of the system, but which is altogether independent of the manner in which that configuration has come about''~\cite{gillispieEdgeObjectivityEssay2016}. Entropy is a way of understanding the arrow of time, the tendency of entropy to increase while energy tends to disperse and reach homogeneity.

Ludwig Boltzmann~\cite{boltzmann1970weitere} took the concept further in his kinetic theory of gases. Boltzmann's proposition is closely tied to the statistical interpretation of thermodynamics. He sought to connect the macroscopic behavior of a system, as described by classical thermodynamics, with the microscopic behavior of its constituent components, such as atoms or molecules. A microstate is the exact configuration or arrangement of positions of entities in the system. A macrostate, on the other hand, represents the macroscopic properties of the system, like temperature, pressure, and energy. A critical concept to think of entropy, particularly in connection with cities, is the definition of \textit{phase space}. This mathematical space represents all possible configurations or positions of components in a system. Boltzmann's approach aims to explain macroscopic properties from the statistical distribution of microstates. Essentially, entropy is related to the number of ways the components in a system can be arranged while still producing the same macroscopic state. 

Entropy is statistical in nature, i.e.\ for a given set of large-scale observable properties, every possible configuration of components that could give those properties is equally likely \cite{odowdTelecollaborationVirtualExchange2018}. Systems with similar macroscopic properties can be characterized by their probability distribution in the phase space. The probability distribution describes the likelihood of finding the system in a given microstate. Basically, a macrostate gives the same or similar average distributions of microstates. The average distribution in the phase space is calculated by considering the probabilities of various microstates. Different distributions can lead to similar averages when they correspond to equivalent macrostates or satisfy certain statistical principles, like the ergodic hypothesis, i.e. over a sufficiently long time, a system will explore all accessible microstates (constrained by the laws of physics) with equal probability~\cite{maxwell2003illustrations,boltzmann1970weitere}. The probability of a microstate is calculated through
\begin{equation}
S=\kappa \ln W
\,,
\end{equation}
where $S$ is the entropy, $\kappa$ is the Boltzmann constant, and $W$ is the number of possible microstates corresponding to a macrostate of the system. Entropy is proportional to the natural logarithm of the number of microstates associated with a particular macrostate. Even though the number of possible microstates may be exceedingly hard to calculate (imagine particles randomly moving within a room), there is no arbitrariness in deciding which states to count in, i.e.\ we equally count all states compatible with the macroscopic parameters~\cite{schwartzLectureEntropy2021}. In the absence of additional information, all microstates consistent with a given macrostate are equally probable, so these random distributions correspond to the same or equivalent microstates and satisfy the ergodic hypothesis. 
The goal is to explain certain macro-properties of the system to a desired degree of accuracy without having to explain the behavior of each individual component of configurations at the micro-level~\cite{wilsonEntropyUrbanRegional1970}. 
A vast number of possibilities of random configurations corresponding to the same average distribution or macrostate contain high entropy. 
In fact, the more microstates corresponding to a macrostate, the higher the entropy. 
The macrostate with the highest number of microstates has maximum entropy.
This reflects the degree of disorder or uncertainty in the system. 

\begin{figure}
    \centering
    \includegraphics[width=\textwidth]{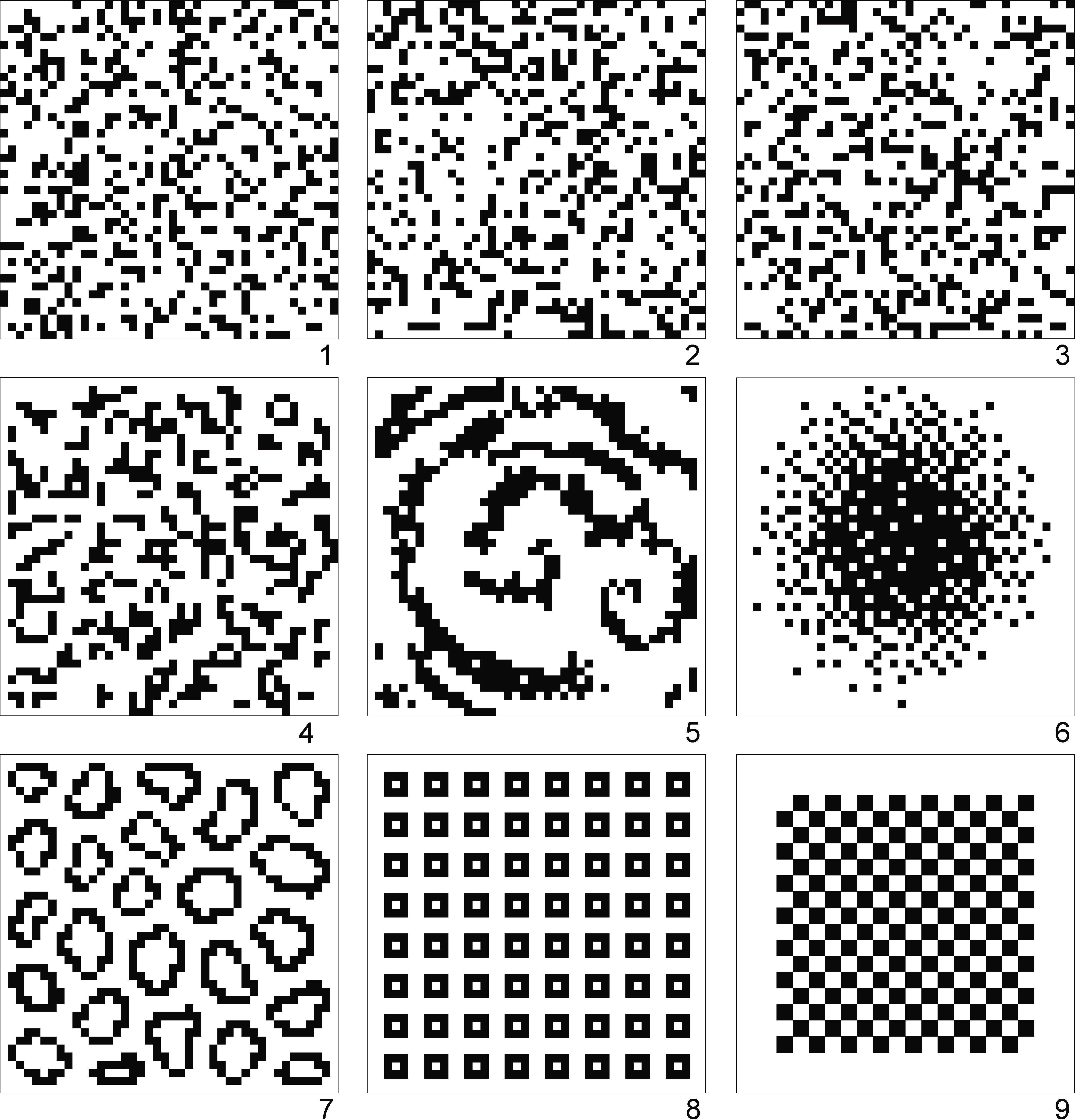}
    \caption{Illustrative configurations in 42$\times$42 grids. All configurations contain 512 black cells, so entropy varies only in relation to distribution. From top left to bottom right: cases 1-3 show different random microstates, which give a similar average distribution and may coarse-grain into the same macrostate. Microstate 4 would give a slightly lower entropy level. Microstates 5-9 show decreasing entropy levels and coarse-grain into different macrostates. Only configurations 6-8 resemble anything close to an urban distribution.
    }
    \label{fig1}
\end{figure}

Imagine a space with two \emph{degrees of freedom} or coordinates to describe the position of cells in a 42x42 cellular grid (Fig.~\ref{fig1}). This space contains 512 black cells whose different distributions represent different microstates or configurations of cells. Every specific configuration has the same probability of occurring, but there are many configurations that have roughly the same macro-properties. The first ones (cases 1-3) show different random microstates which give a similar average distribution and coarse grain into approximately the same macrostate. In fact, the vast majority of possible configurations correspond to such randomly mixed microstates, where cell distributions tend to contain none to low internal correlations. 

In turn, different ordered configurations do not correspond to the same macrostate, like cases 5-9. Internally regular arrangements like spirals, clustered cells dispersing with the distance, checkerboard distributions and cellular rings, either deformed or orthogonal, correspond to rarer microstates in the set of possible microstates, so they seem like drops of order in a sea of disordered states. The larger the system, the less likely it is to find rare configurations like them -- so rare they may never happen. 
At this point, we can only mention that biological structuring or human creation might be needed for such microstates to occur in real systems and become, say, organic bodies or cities (we will come back to this discussion in Sec.~\ref{city_form}).
Even when we consider rare distributions containing internal correlations, such as cases 4-9, only 6-8 resemble urban configurations -- and, in a closer look, only cases 7-8 are capable of generating density through contiguity between cells while keeping permeability between blocks and could evolve into something like a city~\cite{nettoCitiesInformationInteraction2018}. Of course, the very formation of such ring-shaped micro-structures is a highly unlikely event~\cite{hillierSocialLogicSpace1984a,nettoSocialFabricCities2017}.


\subsection{Gibbs entropy} \label{hist_gibbs}
\begin{figure}
    \centering
    \includegraphics[width=\textwidth]{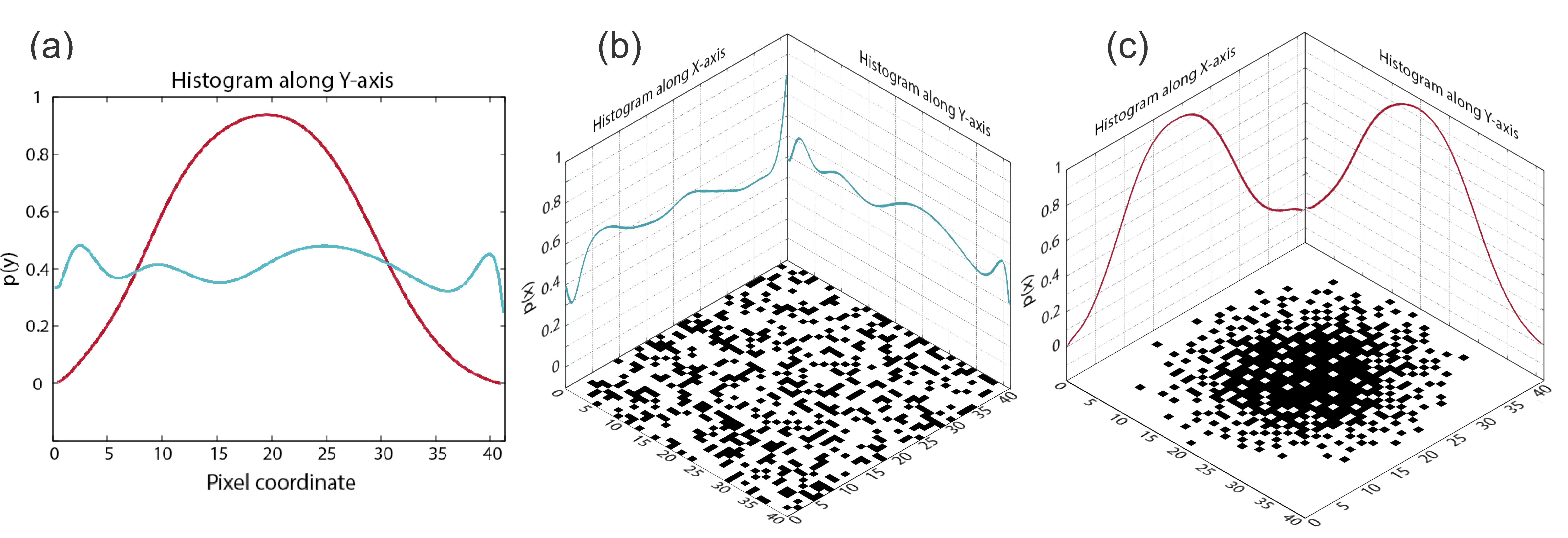}
    \caption{(a) Histogram of the probability of events in a spatial distribution. One scenario with a high level of uncertainty or entropy (blue line) where every possible event (from 1 to 42) had a similar chance of occurring, and another with a low level of entropy (red line), where the probability of occurrence is concentrated around central values. These two probabilities may be illustrated by spatial distributions inspired by Gibbs' probability function. (b) A random distribution where cell positions find no pattern, implying that it would be hard to predict where the next cell would occur; and (c) a clustered distribution of cells with a dropping density according to the distance from the center ($x=0$, $y=0$). In such a distribution, the probability of position of new cells is given by the red curves.}
    \label{fig2}
\end{figure}

Another way to look into entropy in contrast to Boltzmann's~\cite{boltzmann1970weitere} microscopic understanding of the probability of different arrangements, coarse-graining into macrostates is Gibb's~\cite{gibbsElementaryPrinciplesStatistical1902} entropy, a function of the probability distribution over the phase space~\cite{goldstein2020gibbs} (cf.~\cite{liboffGibbsVsShannon1974,gao2019generalized}). Gibbs wished ``to determine how the whole number of systems will be distributed among the various conceivable configurations and velocities at any required time when the distribution has been given for some one time''~\cite{gibbsElementaryPrinciplesStatistical1902}. For a system in equilibrium, entropy is related to the probabilities $P_{i}$ of its microstates
\begin{equation}
S_{G}=-\kappa \sum_{i}P_{i}\ln(P_{i})
\, .
\end{equation}
While the Boltzmann entropy is the entropy for a system in thermodynamical equilibrium, the Gibbs entropy is a generalization of the Boltzmann entropy holding for all systems, i.e.\ including non-equilibrium states. Both are a measure of the microstates available to a system, but the Gibbs entropy does not require the system to be in a single, well-defined macrostate. In equilibrium, all microstates belonging to the macrostate are equally likely, so, for $N$ states, we obtain with $(P_{i}=1/N)$
\begin{eqnarray}
S_{G} &=& -\kappa \sum_{i}\frac{1}{N}\ln\left(\frac{1}{N}\right) \\
    &=& -\kappa N\frac{1}{N}\ln(\frac{1}{N})\\
    &=&-\kappa \ln N
\, ,
\end{eqnarray}
where the latter term is the Boltzmann entropy for a system with \emph{N} micro\-states~\cite{lalinskyWhatConceptualDifference2014}. If the probability distribution is very concentrated at one point in the phase space and very small everywhere else, the entropy state tends to zero~\cite{carroll2020boltzmann}. When the probability of an event, position, or frequency of occurrence is evenly distributed over an area or throughout a space or unit of time, the level of uncertainty in the system dynamics and randomness in its distribution tends to be very high. It is very hard to predict the behavior of a system like this (cf.~\cite{battySpatialEntropyInformation2021}). Differently from the arrangements in Fig.~\ref{fig1}, which illustrated the probability of an arrangement as a whole, we are now dealing with the probability distribution of events or cells occurring within an arrangement (Fig.~\ref{fig2}).

As a measure related to the probability distribution of events within a microstate, the Gibbs entropy may be particularly useful to understand systems with different gradients from mixed to clustered arrangements, such as the locational distribution of land uses (Sec.~\ref{city_diversity}) and the clustering of social groups in spatial segregation (Sec.~\ref{city_segreg}). As we shall see below, cities may not be examples of high entropy distributions but navigate towards low entropy states in varying degrees within the interaction of chaos and order~\cite{crutchfieldWhatLiesOrder2003}.

\subsection{Information theory: Shannon Entropy} \label{hist_shannon}
Shannon’s~\cite{shannon1948} pioneering definition of information was proposed as a measure of uncertainty in the context of the problem of transmission of data, noise, and channel capacity from the point of view of engineering. The fundamental problem of communication is that of reproducing at one point a message selected at another point, either exactly or approximately. If the number of messages in a set is finite, entropy is a measure of the distribution of probabilities that certain messages will occur. Similarly to Gibbs and Boltzmann, it defines entropy in terms of probabilities $P_{x}$ of events. It quantifies the efficiency of a system, such as a code or language, in transmitting information. It is defined as the logarithm of the number of distinct messages that can be conveyed by selecting from the same set of symbols.
\begin{equation}
H = -\sum P_{x} \log_{2} (P_{x})
\, .
\end{equation}
This measurement gives the extent of initial uncertainty that can be resolved by a single message. It is a fine-grained approach to configurations such as the frequencies in which certain signs appear together in strings of symbols. Higher frequencies of certain co-occurrences amount to redundancies that reduce surprise (information) and uncertainty (entropy) in language. Shannon~\cite{shannon1948} illustrates this through a series of experiments, from random arrangements of symbols to restrictions to randomness based on the frequency of symbols found in words and sentences in English. In a zero-order approximation, independent symbols are listed randomly based on the same probability of occurrence.
\begin{quote}
XFOML RXKHRJFFJUJ ZLPWCFWKCYJ FFJEYVKCQSGHYD QP\-AAMKBZAACIBZLHJQD.
\end{quote}
In a first-order approximation, independent symbols have the same probability of appearing that they have in English.
\begin{quote}
OCRO HLI RGWR NMIELWIS EU LL NBNESEBYA TH EEI ALHENHTTPA OOBTTVA NAH BRL.
\end{quote}
In a second-order approximation, we have the frequency with which a symbol follows another in pairs in English.
\begin{quote}
ON IE ANTSOUTINYS ARE T INCTORE ST BE S DEAMY ACHIN D ILONASIVE 
TUCOOWE AT TEASONARE FUSO TIZIN ANDY TOBE SEACE CTISBE.
\end{quote}
In a third-order approximation, symbols are randomly sequenced according to the likelihood of association with others in trigrams.
\begin{quote}
IN NO IST LAT WHEY CRATICT FROURE BIRS GROCID PONDENOME OF DEMONSTURES 
OF THE REPTAGIN IS REGOACTIONA OF CRE.
\end{quote}
A first-order approximation to words shows symbols chosen independently from each other but with the same probability of appearing that they have in sentences in English.
\begin{quote}
REPRESENTING AND SPEEDILY IS AN GOOD APT OR COME CAN DIFFERENT NATURAL HERE HE THE A IN CAME THE TO OF TO EXPERT GRAY COME TO FURNISHES THE LINE MESSAGE HAD BE THESE.
\end{quote}
In a second-order word approximation, Shannon comprises the probability that a word is followed by another specific word, with no further structure included,
\begin{quote}
THE HEAD AND IN FRONTAL ATTACK ON AN ENGLISH WRITER THAT THE CHARACTER OF THIS POINT IS THEREFORE ANOTHER METHOD FOR THE LETTERS THAT THE TIME OF WHO EVER TOLD THE PROBLEM FOR AN UNEXPECTED.
\end{quote}
A nearly intelligible message emerges even without syntactic rules or decoding the meanings of words. 

Differences in frequency and sequence in symbols and words produce differences in the probability of occurrence and an increased probability of certain combinations. This means that the message components become more predictable and intelligible. The reduction of entropy allows communication~\cite{nettoSocialInteractionCity2017}. This is also the limit to Shannon's approach without semantic information. Most importantly, as we have seen above, the entropy evaluation of a given probability distribution is a unique and unambiguous means of quantifying the intuitive difference between a broad distribution and a sharply peaked one~\cite{brigattiEntropyHierarchicalClustering2021,jaynesInformationTheoryStatistical1957}. This notion can be naturally related to a general idea of uncertainty and randomness, where larger entropies correspond to broader distributions closer to equiprobability and randomness. Of course, Shannon proposed this formulation in a context where his problem was to understand sequences of information-carrying symbols~\cite{shannon1948}. Shannon’s entropy quantifies the uncertainty associated with predicting a letter which follows a well-known portion of a string. Analogously, we shall see that increased frequencies of associations beyond one-dimensional strings, namely in three-dimensional arrangements of spatial entities like buildings, will create consistencies, correlations and structures in spatial formations fundamental to cities and their workings -- and will set them apart from random distributions (see Sec.~\ref{city_form}).

\subsection{Entropy in biological systems} \label{hist_biol}
Another rich source of influence on the study of entropy in cities stems from the use of the concept in biology. Interpretations of the concept from Physics were used to understand life -- from the thermodynamic conditions for the origin of life and the difference between living and non-living structures to the emergence of organized biological systems and their recursive adaptation in the face of environmental fluctuations and change. The second law of thermodynamics asserts that a closed system will inevitably experience an increase in entropy. However, living systems seem to challenge this law by maintaining low entropy. According to Prigogine~\cite{prigogine1945moderation}, such systems will take on the configuration that minimizes its rate of entropy production~\cite{ulanowicz1987life}. Boltzmann himself first highlighted this concept, framing life as a struggle against entropy~\cite{odowdTelecollaborationVirtualExchange2018}. The thermodynamic understanding of life in the face of entropy posits that living systems diminish their internal entropy while elevating the entropy of their surroundings. They require energy to synthesise and maintain their highly ordered structures. This energy is often obtained through processes like metabolism, where complex molecules with a high degree of order (e.g.\ glucose) are broken down to release energy, also releasing simpler end products (e.g.\ carbon dioxide and water) and heat as a byproduct, increasing the disorder of the molecules in the environment~\cite{ulanowicz1987life}. This idea was closely interpreted in the study of urban metabolism and its entropic environmental effects (see Sec.~\ref{city_sustainability}).

Erwin Schr\"odinger~\cite{schrodingerWhatLifePhysical1944} further explored this idea depicting life as a process that thrives on negative entropy. This is the case in the emergence of \emph{order in the form of organic structures}, from molecules to cells to whole bodies. Living systems are characterized by highly specific and non-random structures. Atoms arrange themselves in organic molecules through physical and chemical combinatorial properties of atoms like carbon, which allow them to combine with other elements. Such far from random combinations form molecules like DNA. Schr\"odinger called these structures `aperiodic crystals'~\cite{schrodingerWhatLifePhysical1944,portugaliSchrodingerWhatLife2023}. As microstates, such ordered configurations are statistically rare events in the space of possible configurations -- except in the realm of biological systems, where combinations can achieve high frequency because such systems are able to synthesise and replicate themselves. DNA is an emblematic example. The molecular bases of DNA form four possible configurations connecting the double helix and amount to information-carrying sequences needed for coding other molecules and cellular structures. The micro-structures of DNA guide the \emph{morphogenesis} of large structures emerging from the space of possible configurations, or what evolutionary biologists call \emph{morphospace}~\cite{raup1966geometric,buddMorphospace2021}. Exposed to random fluctuations in their external milieu, such self-assembling systems restrict themselves to occupying a limited number of states. Distilling structural regularities~\cite{fristonFreeEnergyPrinciple2012}, they can form specialized subsystems like muscles and bones or circulatory and neural networks. In this \emph{autopoietic formation of structures in space}~\cite{maturanaAutopoiesisCognitionRealization1991} \emph{across scales}, subsystems are connected and integrated into the functioning whole of the organism. Organisms are arrangements operating in coordination at different scales -- a principle that came to pervade even the understanding of societies, from Durkheim~\cite{durkheimRulesSociologicalMethod1895} to Luhmann~\cite{luhmannSocialSystems1995}, and cities as systems of specialized, mutually-dependent subsystems (see Sec.~\ref{city_self-org}). Of course, all these steps in the emergence of structures from smaller, nested structures are rare events in the horizon of possible events - even though actual organisms occupy a comparatively even smaller number of configurations in morphospace, in the non-ergodic universe above the level of atoms~\cite{kauffman2019world}.

Uses of the concept of entropy in biology in ways appealing to urban metaphors were also applied to understand how living systems \emph{survive}, how they respond to their environment and are able to improve and adapt to changing conditions. They have the ability to \emph{resist a natural tendency to disorder}~\cite{fristonFreeEnergyPrinciple2012}, i.e.\ their own degradation and decay as they metabolise and self-repair. Such resistance includes the production of specialized subsystems geared to protect and repair their structures, from immunological systems, as organisms are not closed systems in equilibrium, to the internal cleaning of entropic outputs of their metabolism. Such an ability is addressed as `delayed entropy’ by Schr\"odinger. ``How would we express the marvellous faculty of a living organism, by which it delays the decay into thermodynamic equilibrium (death)?''~\cite{schrodingerWhatLifePhysical1944}. Life is a property of a system to temporarily delay the process that leads the system toward maximum entropy~\cite{portugaliSchrodingerWhatLife2023}. Such biological interpretations of entropy influenced approaches to cities as metabolic systems drawing energy and self-structuring in the face of environmental fluctuations and randomness in their own morphogenesis while spilling effects of their growth and workings over their environment. Cities can self-repair and improve in the face of entropic forces -- properties explored in views of urban self-organization and sustainability (Sec.~\ref{city_self-org} and \ref{city_sustainability}).

\subsection{Entropy in social systems} \label{hist_social}
The concept of entropy has been explored in relation to societies as systems co-evolving with cities. Several theorists, within the framework of Social Entropy Theory (SET) originating in the 1960s, have specifically employed Shannon's information measure~\cite{buckleySociologyModernSystems1967,mcfarlandMeasuringPermeabilityOccupational1969}. Charv\'at et al.~\cite{charvatSystemTheoryDependence1973} introduced concepts like `entropy of behavior' to assess homogeneity and interdependence. Horan~\cite{horanInformationTheoreticMeasuresAnalysis1975} developed a measure of proportional reduction of uncertainty. Beyond internal entropy accumulation, threats to the social system also arise from the external environment. Entropy is viewed as the antithesis of information, controlled by system boundaries~\cite{klappOpeningClosingOpen1975}. Galtung~\cite{galtungEntropyGeneralTheory1975} analyzed entropy at micro- and macroscopic levels, identifying actor entropy and interaction entropy. Strong forces propel social systems into an oscillation between low- and high-entropy states. Action outcomes are not random, as agents operate within generalized roles, norms, and goals. Social norms, rules, culture, and language serve as constraints preventing maximum entropy. In turn, Bailey~\cite{baileySociologicalEntropyTheory1983,baileySocialEntropyTheory1990} looked into how social systems increase organizational complexity and decrease internal entropy over time. The replication of recurring actions leads to order, resulting in entropy less than maximum. Removing action constraints could lead to increased entropy levels.

An insightful approach deals with counterfactuals and quasi-ergodic possibilities as entropic forces opened up by and surrounding people’s daily actions. Luhmann~\cite{luhmannSocialSystems1995} viewed societies as networks of interconnected subsystems where people face the challenge of selecting actions from an increasing array of possibilities~\cite{millerLivingSystems1995}. In societies with growing information and agency, the complexity of possible interactions rises exponentially, leading to difficulties in selection. There is an increase in an unstructured type of complexity, an entropic informational complexity that may lead to organizational loss. The concept of \emph{unstructured complexity} unraveled by possibilities that pose challenges to the selection and actualization of actions illuminates counterfactual risks that social systems recursively encounter.

Other approaches consider societies as collections of agents within geographic boundaries~\cite{millerLivingSystems1995}. Bailey~\cite{baileySociologyNewSystems1994,baileySociocyberneticsSocialEntropy2006} explored entropy in relation to city and population size. Social cybernetics concepts, such as context-dependency and agent self-reflexivity, were applied to issues like the division of labor, communication challenges, adaptation in complex societies, information's role in decision-making, and entropy reduction~\cite{geyerChallengeSociocybernetics1995,geyerMarchSelfReference2002}. However, these conceptualizations often lack spatial depth, treating space as a background rather than an integral part of entropy's resolution. Previously marginalized in such approaches, recently cities began to be seen as complex and active environments influencing and co-evolving with social systems~\cite{nettoSocialInteractionCity2017,nettoCitiesInformationInteraction2018} (see Sec.~\ref{city_self-org}).

We shall now see how these and other aspects of entropy have been interpreted and influencing the understanding of cities as complex systems.

\section{Exploring the concept of entropy to understand cities: domains, measures and applications} \label{cities_3}
The use of the concept of entropy in urban studies draws upon thermodynamics, statistical physics, and information theory, along with their biological and sociological interpretations. As hinted above, applying the concept of entropy to cities may provide insights into how they emerged in different regions of the world and evolved into complex systems of an enormous diversity of shapes and forms. As interest in the concept of entropy grew in recent decades in urban science, several domains and properties have been explored.

\subsection{Entropy-maximizing urban systems and spatial entropy} \label{city_maxim}
The introduction of entropy concepts in relation to cities originated in the late 1950s. (i) Drawing analogies to statistical thermodynamics, it was motivated by concerns about how urban systems evolve toward a steady state (see~\cite{battySpatialEntropyInformation2021}. Researchers in the early 1960s (such as~\cite{leopoldConceptEntropyLandscape1962,curry1964}) related entropy increase in closed systems to the dynamics of landscape ecologies and human systems. (ii) Other efforts focused on quantifying entropy as a measure of order to assess spatial object concentration or dispersion. (iii) A third line of approach culminates in the method of entropy maximizing focused on identifying constraints to the amount of entropy in an urban system. This method derives models of the distribution of energy in the system, introduced as a means to study the dynamics of urban systems by Wilson~\cite{wilsonEntropyUrbanRegional1970}. The concept was used to approach the most likely macro-distribution for micro-level interactions in a spatial economic system, namely through spatial interaction models able to capture the emergence of equilibrium solutions~\cite{reggianiHandbookEntropyComplexity2021}, in dynamics applied to transport flows and retailing, relating urban disaggregation to utilities and disutilities costs~\cite{wilsonEntropyUrbanRegional1970}.
The entropy-maximizing paradigm was used to derive model formulations for spatial interactions and urban distributions, micro-economic behavior and input-output analysis~\cite{battenSpatialAnalysisInteracting1982}. Wilson~\cite{wilsonEntropyUrbanRegional1970} originally pursued an analogy with the statistical mechanics of entropy to find a way to reproduce the then-standard gravity model 
and a wide range of applications involving flows and locational structures related to the cost of transports and retail location models. Routes, flows and a disaggregation process emanate from the urban core. 

The approach involves economic cost-benefit analysis in transport planning, studied through micro-economics of discrete choices~\cite{wilsonEntropyUrbanRegional2010}. The concept is used to generate interaction patterns of the journey to work subsystem, while indices of information theory entropy are used to unravel the nature and structure of these flows or movements of people~\cite{ayeniCitySystemUse1976}. For any flow pattern, a complete dispersion of inputs and outputs and, hence, economic linkages means maximum uncertainty in tracing the course of each flow. Furthermore, where there is a maximum concentration of inputs, there is little uncertainty in finding a place in the system for a randomly selected flow~\cite{ayeniCitySystemUse1976}. The most probable state of the elements of the system is achieved when its entropy is maximized subject to the constraints on the microstate of the system~\cite{wilsonEntropyUrbanRegional1970}. 

Harris and Wilson~\cite{harrisEquilibriumValuesDynamics1978} extended the entropy-maximizing framework, embedding slow dynamics into spatial interactions. In a series of works in the 1970s, Batty~\cite{battySpatialEntropy1974a,battyEntropySpatialAggregation1976} explored a definition of spatial entropy as an evolving entropy-maximizing model introduced by Wilson, applied to the spatial aggregation problems, suggesting principles to partitioning spatial systems and uses of spatial entropy to cut off urban boundary to explore questions about optimal geometry of spatial systems. From then on, spatial interactions became widely related to Boltzmann's statistics methods, associating probabilities and maximization models of urban systems~\cite{wilsonEntropyUrbanRegional1970}. Entropy was linked to gravity models, rank-size rule \textbf{(see Chapter X)} and discrete choice models as a measure of interaction, uncertainty, diversity etc.~\cite{reggianiHandbookEntropyComplexity2021}. For instance, Leibovici et al.~\cite{leibovici2009defining,leibovici2014local} consider as spatial information the distribution of co-occurrences between two or more observations with varying vicinities for the exploration of spatial patterns at different scales.

\subsection{Entropy, information and complexity in cities} \label{city_shannon}
Shannon's entropy certainly offers multiple domains encompassing order-disorder, certainty-uncertainty, minimum to maximum entropy, and zero to complete information~\cite{battyEntropyComplexitySpatial2014,battySpatialEntropyInformation2021}. His definition blurred the boundaries between entropy, uncertainty, and information. Weaver~\cite{weaver1953recent} himself pointed out that the idea that ``greater uncertainty and greater information go hand in hand'' is deeply counter-intuitive. In fact, divergences between the statistical mechanic sense and information theory-based approaches on the relationship between entropy, uncertainty, and information were already felt in urban studies in the 1970s~\cite{battySpatialEntropy1974a,ayeniCitySystemUse1976}. More recently, the concepts of entropy and information have been more related through the appropriate lens of complexity. Batty et al.~\cite{battySpatialEntropyInformation2021} propose a measure of complexity based on Shannon’s information to grasp cities as they vary in scale, size, and spatial distribution of population. They interpret complexity in opposition to the idea of intricate systems able to hold order together in the face of unpredictable environments. 

Batty chooses the interpretation that when something is entirely ordered, hence completely predictable, it is no longer complex. Of course, this conceptual choice renders the approach compatible with Shannon's view of entropy and information. The next step is to bring a view of the city as a distribution of events -- namely, locations in relation to population size and area. Complexity is admitted by considering growth in the number of events (locations) and changes in their distribution. They adapt Shannon's measure partitioning information into two components, (i) a spatial entropy dealing with the distribution of information and (ii) an information density that deals with size. The idea is that the complexity of a system is inherently tied to the number of events it encompasses. Shannon's measure ensures that there is a trade-off between the distribution of information, the number of events, and their density as the urban system changes in size and scale, and the distribution evolves from highly ordered to less ordered, or vice versa.

Various methods in landscape ecology have explored adaptations of Shannon entropy and information-theoretical metrics to quantify and group similar two-dimensional spatial patterns. Some of these approaches aim for a universal classification of configuration types, arranged linearly based on increasing values~\cite{claramuntSpatiotemporalFormEntropy2012,altieriNewApproachSpatial2018,nowosadStochasticEmpiricallyInformed2019}. In the context of cognitive studies, efforts to extract task-relevant information from the built environment, other spatial information approaches have used measures of entropy, distribution of spatial co-occurrences, or information density~\cite[e.g.]{rosenholtzMeasuringVisualClutter2007,woodruffConstantInformationDensity1998}. Additionally, Haken and Portugali~\cite{hakenportugali2003,haken2015information} have explored the embodiment of information in the built environment, quantifying Shannon's information in connection with Haken's synergetic qualitative approach to semantic information. Their work empirically assesses how basic cellular arrangements and categorizations of building facades convey varying amounts of information. 

\subsection{Entropy and urban form} \label{city_form}
There is a distinction between analyses of distributions of densities, locations, or activities and approaches geared to reveal correlations within distributions of urban form involving spatially well-defined entities such as buildings and streets. The latter aims to grasp structures at different scales, from detailed arrangements of buildings or streets at fine resolutions to large-scale structures in spatial morphologies. Approaches to entropy in urban form range from applications as measures of diversity as a proxy to disorder in street networks to explorations to capture information signatures in the built form distributions of cities in different world regions.

\subsubsection{Entropy in street networks} \label{city_form_networks}
\begin{figure}
    \centering
    \includegraphics[width=\textwidth]{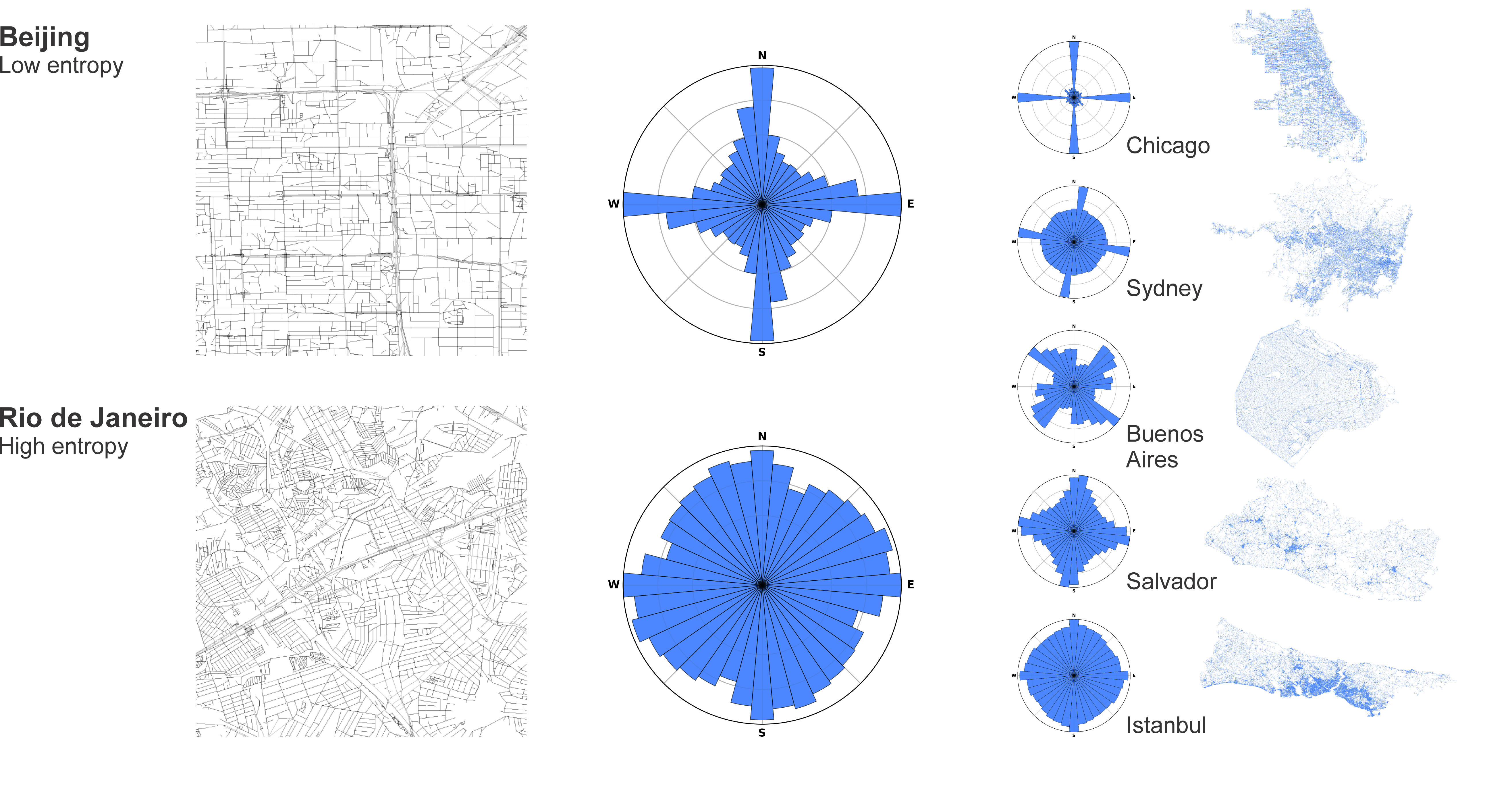}
    \caption{Entropy measures can be applied to analyze variation in street orientations illustrated by polar histograms, e.g. Beijing and Rio de Janeiro (left and center), and Chicago, Sydney, Buenos Aires, Salvador and Istanbul (right).}
    \label{fig3}
\end{figure}

\paragraph{}
The understanding of cities as topological structures was systematically introduced by Christopher Alexander in the mid-1960s~\cite{alexanderNotesSynthesisForm1964,alexanderCityNotTree1965}. This graph-theoretical view ushered into approaches to configurations of buildings \cite[e.g.]{steadmanArchitecturalMorphologyIntroduction1983,hillierSocialLogicSpace1984a} and cities~\cite{hillierSpaceSyntaxDifferent1983}. The space syntax approach offered the first topological analysis of cities as street networks to address the systemic structures and workings of cities as part of the social dynamics of people's movement and encounter systems. This line of research was popularized as (and reduced to) a purely morphological approach as `street network analysis' in the 2000s \cite[e.g.]{jiangTopologicalAnalysisUrban2004,claramuntSpatiotemporalFormEntropy2012,crucittiCentralityMeasuresSpatial2006,portaNetworkAnalysisUrban2006,barthelemyModelingUrbanStreet2008}. Gudmundsson and Mohajeri~\cite{gudmundssonEntropyOrderUrban2013} explored street networks to introduce a method based on Gibbs/Shannon entropy to measure angular and length variation between streets in 41 British cities: 
\begin{equation}
S=-\kappa\sum_{i=1}^{t}P_{i}\ln(P_{i})
\, ,
\end{equation}
where $S$ is entropy, $\kappa$ is a positive constant, $t$ is the number of bins with nonzero probabilities of streets, and $P_{i}$ is the probability of streets falling in the $i$-th bin. Entropy values vary between zero (all the streets occupy a single bin) to the maximum value (a uniform distribution or lengths so that all bins have the same lengths or heights (Fig.~\ref{fig3}, right). Findings show that entropy increases with street average length, distance from the city center and as cities group over time. 
Boeing~\cite{boeingUrbanSpatialOrder2019} further developed the approach into an orientation-order indicator to quantify the extent to which a street network follows the spatial ordering logic of a single grid, and other indicators like street circuity applied to 100 cities worldwide and show consistent regional differences in values. Coutron et al.~\cite{coutrotEntropyCityStreet2022} explored such properties in a cognitive study to measure spatial navigation ability in more than 397,000 subjects from 38 countries through cognitive tasks embedded in a video game. Their findings revealed that people demonstrated better navigation performance in environments topologically resembling their upbringing. Growing up in cities with low street network entropy, such as Beijing (Fig.~\ref{fig3}, left), resulted in better performance in digital environments with a regular layout. Conversely, those growing up outside cities or in cities with higher street network entropy, like Rio de Janeiro, had better performance in less ordered digital environments.

\begin{figure}[t]
    \centering
    \includegraphics[width=\textwidth]{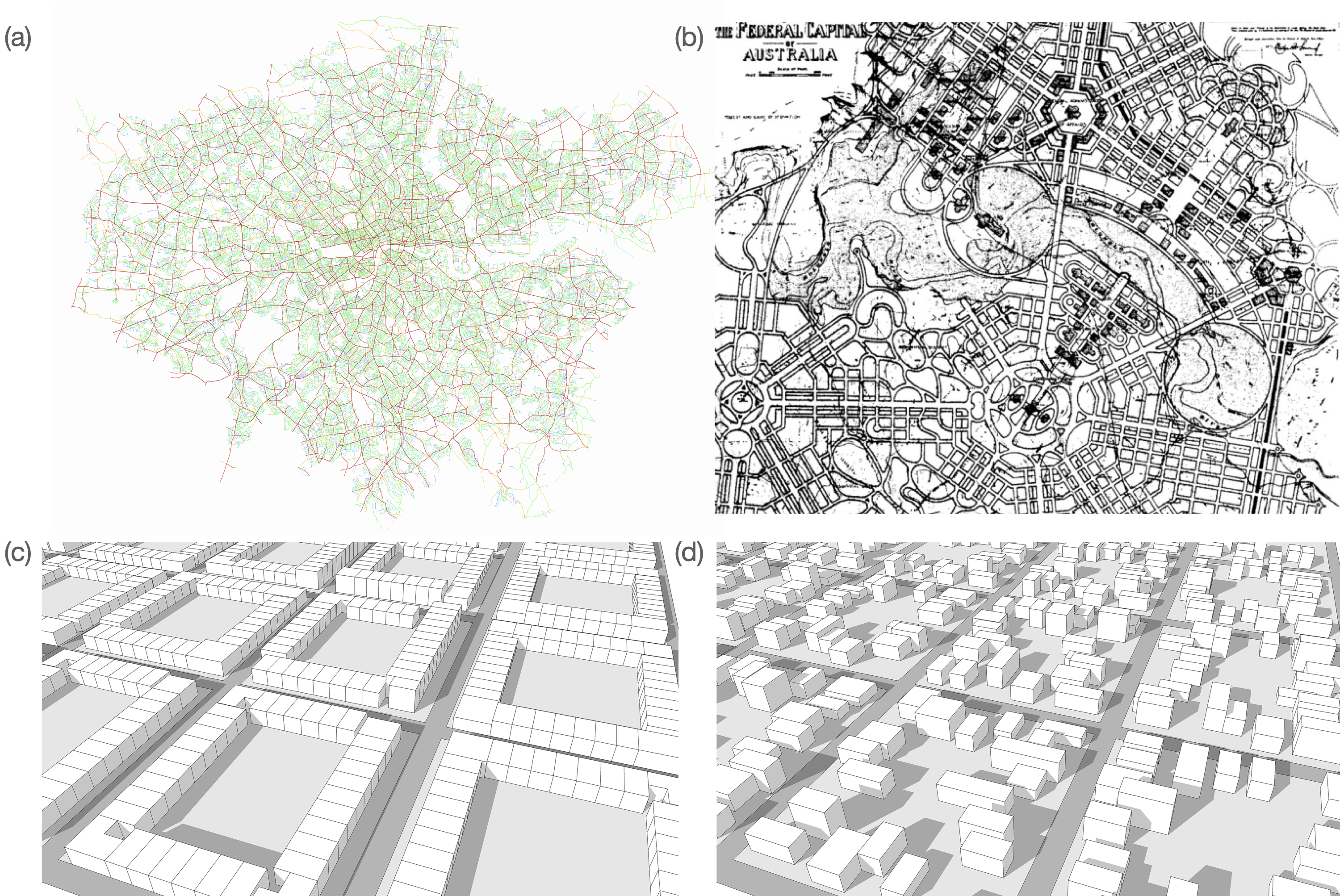}
    \caption{High angular variation can be found in (a) the large-scale structures of `organic' cities like London, UK, topologically identified through space syntax methods \cite{SpaceSyntaxOpenMapping}, and in (b) perfectly ordered radial street networks like the central area of Canberra, Australia, as designed by Walter Burley Griffin in 1911. The same street network in (c) and (d) can support very different built-form systems, e.g. ordered (bottom left) and random distributions of buildings (bottom right)~\cite{nettoUrbanFormInformation2023}.}
    \label{fig4}
\end{figure}

Note that Shannon entropy is more than a general diversity index. It is a well-defined measure of information that quantifies the degree of surprise the source of a sequence causes on the observer~\cite{shannon1948,brigattiEntropyHierarchicalClustering2021}. 
Therefore, this interesting method focusing on entropy as a measure of variation in street orientations and lengths seems yet to be fully explored to uncover spatial information patterns. 
The entropy measure of the distribution of local traits like crossing angles or street orientations and lengths does not necessarily capture the global degree of order/disorder of street networks as long-range correlations. The apparent randomness at local scales frequently amounts to structure if looked at a larger scale (Fig.~\ref{fig4}a). High angular variation may be found in ordered cities, of which concentric radial street networks are the simplest case (Fig.~\ref{fig4}b). Furthermore, entropy-based diversity approaches applied to street networks can ignore the complexities of urban form latent in the distribution of buildings and the discrepancies between the order in street networks and in built form systems. Of course, cities can be spatially disordered even if their street networks are perfectly ordered: they may have low entropy in street orientation yet highly disordered morphologies~\cite{nettoUrbanFormInformation2023} (Fig.~\ref{fig4}d).

\newpage

\subsubsection{Entropy in built form systems} \label{city_form_built}
\paragraph{}
A quite different situation corresponds to the built environment as a systemic whole, not selecting just a single local trait for characterizing it. Entropy measures can be applied to understand the ultimate feature defining urban form structures, i.e.\ buildings producing built-form configurations. Morphogenesis is a concept that originated in biology. In the context of cities, it addresses the process by which the form, structure, and organization arise during the development of cities.
The emergence of structures in cities in the face of the uncertainty of possible morphogenetic paths is a crucial, if underestimated problem in urban science. The process may be approached as restrictions on randomness \cite{hillierSocialLogicSpace1984a}, echoing Shannon's~\cite{shannon1948} view of information as the amount of freedom of choices governed by probabilities. This is an open question in urban research. We may hypothesize that patterns emerge in the course of local societies actually solving problems of aggregating buildings into complexes and settlements. Such spatial solutions and formations seem to evolve in time through trial and error, conscious and unconscious spatial choices, contingencies and path dependence, among other potential factors~\cite{nettoSocialFabricCities2017,alexanderNotesSynthesisForm1964,hillierSocialLogicSpace1984a}. The effect is the reduction of the uncertainty manifested as the increase in the probability of certain spatial distributions of buildings and open spaces. Higher frequencies bring regularities and internal correlations at different scales which, ultimately, may amount to large-scale structures in spatial formations. 

\begin{figure}
    \centering
    \includegraphics[width=\textwidth]{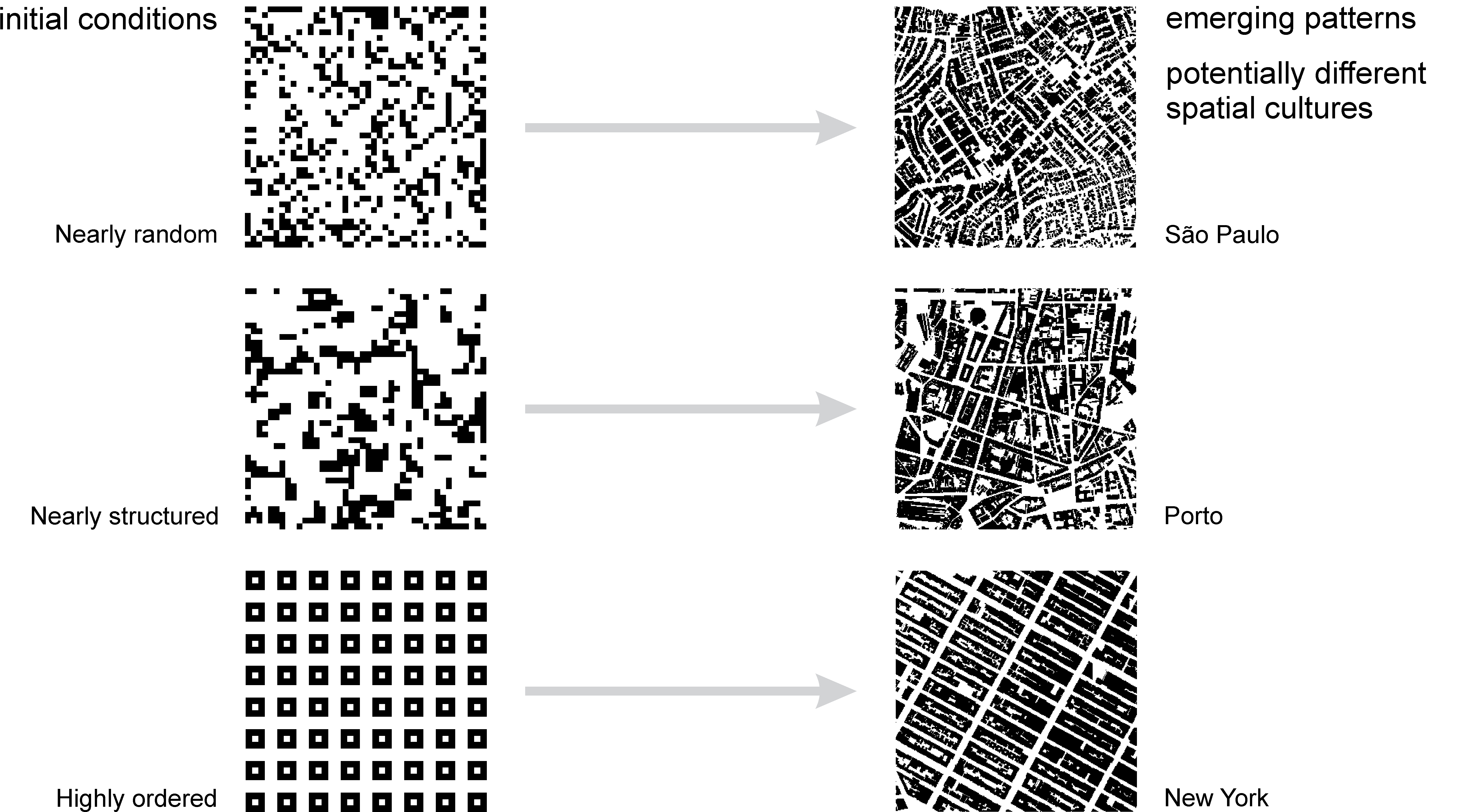}
    \caption{Illustrations of possible archetypal morphogenetic paths. Initial conditions with different entropic levels (left) might evolve into distinct urban patterns and specific spatial cultures (right): São Paulo, Brazil; Porto, Portugal; New York, USA.}
    \label{fig5}
\end{figure}

Of course, these formations may take different morphogenetic trajectories across morphospace. Trajectories might be contingent and relate to local environmental conditions and societal codes -- a plethora of factors that amount to cultural specificities. Here, the amount of randomness in initial morphogenetic conditions might matter. They might be related to culturally embedded choices bringing restrictions to the random creation of urban form, leading to different outputs which, over time, consolidate as patterns and, through path dependence, into distinct urban spatial cultures~\cite{nettoUrbanFormInformation2023,hillier1989}. Some spatial cultures might favor order and rules guiding additions to growing settlements. Others might be more relaxed when engaging in morphogenesis, allowing purely bottom-up processes to take over (Fig.~\ref{fig5}).

\begin{figure}
    \centering
    \includegraphics[width=\textwidth]{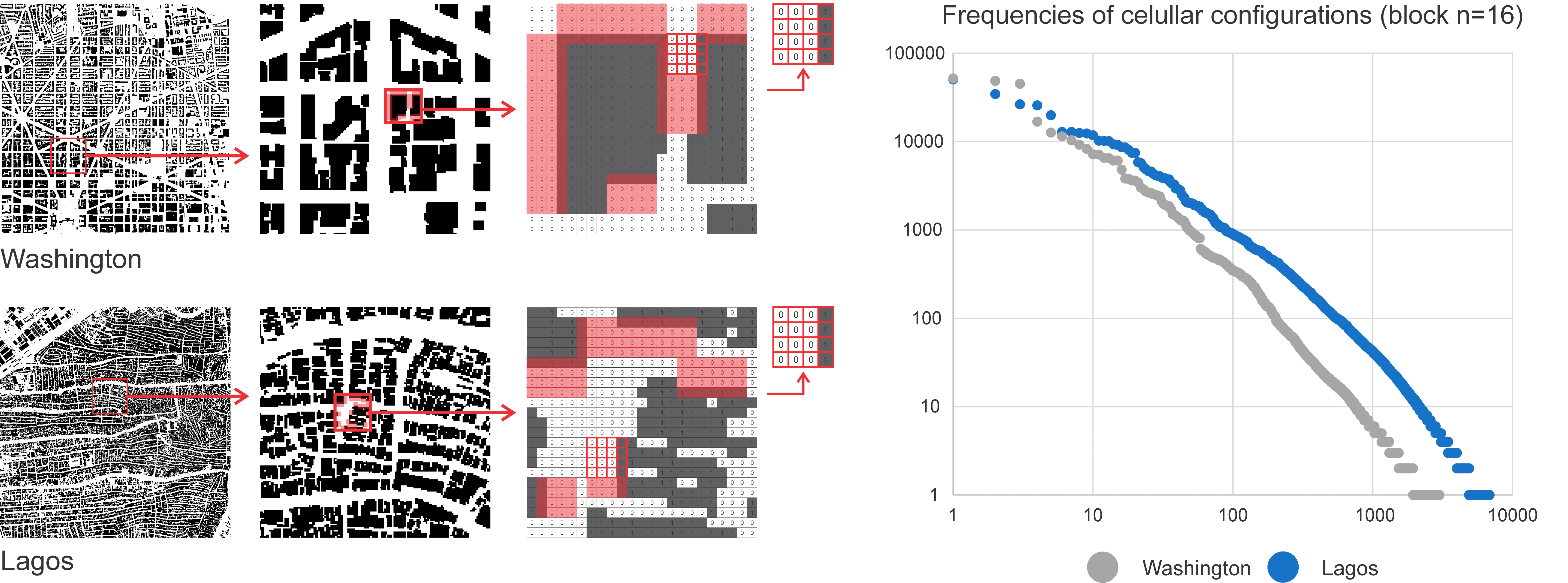}
    \caption{Frequencies of cellular combinations in different urban contexts. The algorithm scans 9\,km$^2$ areas in built form maps counting frequencies of cellular configurations for different block sizes (from $n=1$ to $n=20$). Amplified areas in red show a specific configuration (a cell block with $n=16$) and where they are found. The total number of configurations $n=16$ found in Washington (grey line) is 2,989 and 6,825 in Lagos (blue line) out of 65,534 possible combinations.}
    \label{fig6}
\end{figure}

Recent works \cite{brigattiEntropyHierarchicalClustering2021,nettoUrbanFormInformation2023} have explored built form configurations to identify levels of entropy and disorder as information signatures of possible cultural traits. The approach uses Nolli maps of building footprints (representations of buildings and city blocks by black polygons or cells, and the open spaces of streets, squares and courtyards by white polygons or cells) as representations of the complete urban form of cities.
Nolli maps are converted into cellular arrangements, a matrix of binary values, i.e.\ an information-carrying image where the symbols correspond to built or non-built spaces. The method allows the identification and counting of frequencies of cellular combinations in different urban contexts. Fig.~\ref{fig6} shows the 1,000 most frequent cellular configurations from all possible combinations for a cell block with $n=4\times 4$, excluding configurations completely built or completely open, in 9\,km$^{2}$ areas of two cities, i.e.\ Washington (US) and Lagos (Nigeria). The total number of configurations found in Washington is 2,989; in Lagos, 6,825 (out of 65,534 possible cellular combinations). These distributions and numbers suggest substantial differences in how these distinct spatial cultures materialize morphogenetic possibilities and emphases on regularity and variety in urban form.

\begin{figure}
    \centering
    \includegraphics[width=\textwidth]{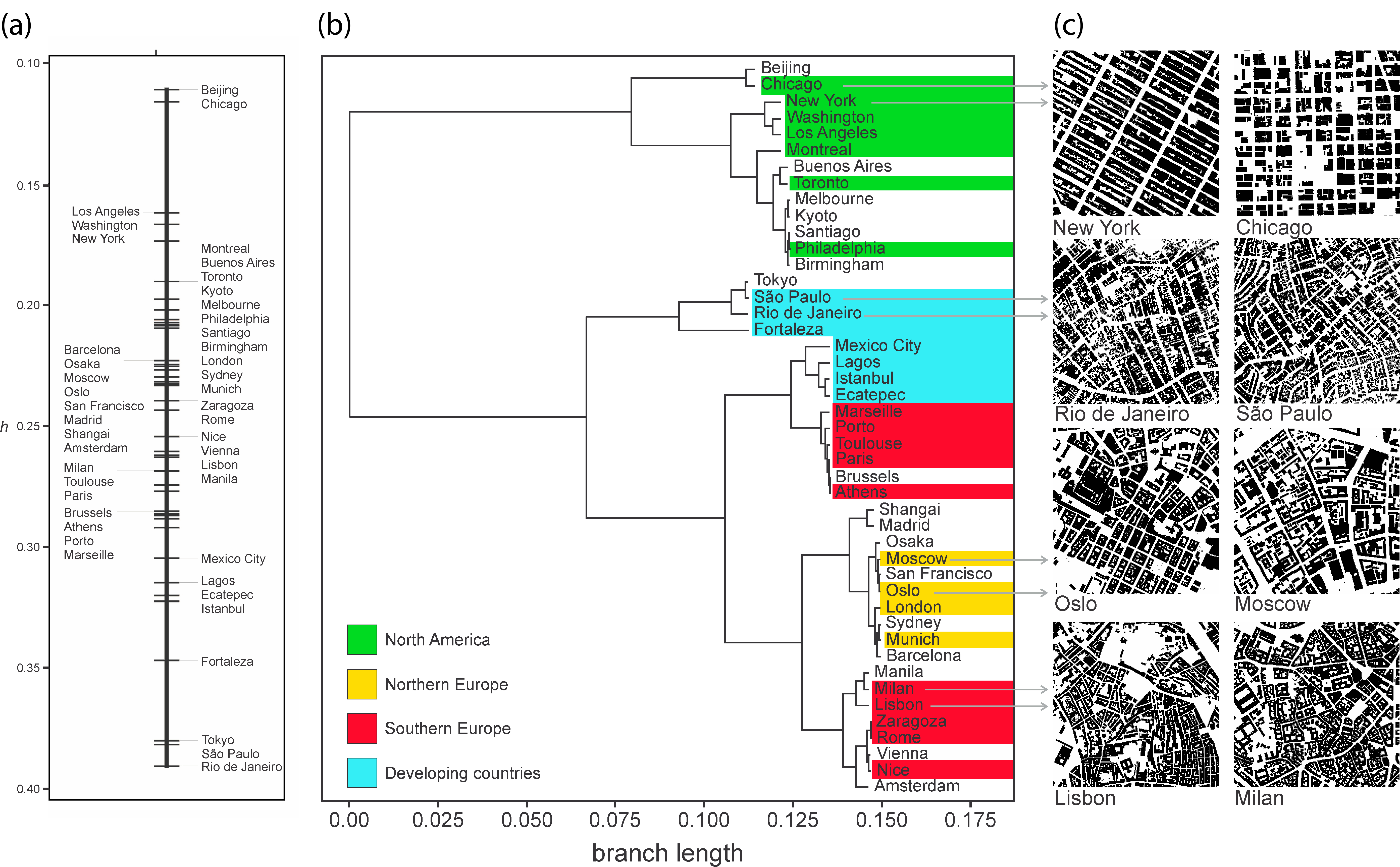}
    \caption{Entropy levels as a proxy to differences in spatial cultures. (a) Estimated values of $h$ for the 45 cities under analysis (see~\cite{nettoUrbanFormInformation2023}). (b) Dendrogram of North American, European, Asian, Oceanian, African, and South American cities under analysis. Colors highlight the relative position of cities regarding world regions. (c) Sub-sections (1\,km$^2$) show the morphologies of some of these cities, illustrating the method’s potential to grasp distinct spatial cultures.
    }
    \label{fig7}
\end{figure}

The method scans urban sections considered satisfactory to represent the spatial characteristics of different cities and to grasp correlations at sufficiently large scales, a necessary condition for correctly estimating entropy. Shannon entropy is applied to measure the amount of randomness and statistical complexity of the global system. To go beyond the local properties of the map, the distribution probability of the state of the entire image must be quantified. As the urban space is represented by matrices of 0 and 1 (open space cells and built form cells, respectively), this procedure theoretically corresponds to measuring the Shannon entropy $H$ of a two-dimensional binary cellular sequence. The method defines the block entropy of order $n$ through
\begin{equation}
H_{n}=-\sum_{k} P_{n} (\kappa) log_{2} [(P_{n})(\kappa)]
\, ,
\end{equation}
where a block of size $n$ is a segment of $n$ cells selected from the sequence under analysis, and the sum runs over all the $k$ possible $n$-blocks. Two blocks with the same number of 0 and 1 but with a different spatial combination of them correspond to different configurations. The method counts the number of times each configuration $k$ is found in the area, accounting for actual built form cellular configurations and not merely the density of cells (occupation density). Scanning cellular blocks from $n-1$ to $n=20$ allows for averaging entropy values across larger blocks of cells. The method considers features of longer-range order as it computes frequencies and identifies the same configurations in different locations, even quite far from each other, as correlations at a distance. 
Physical arrangements marked by higher levels of randomness and higher entropy levels are characterized by greater unpredictability. 
In contrast, patterns and regularities in urban structures correspond to lower entropy, which means a higher predictability. 
Results suggest some form of cultural idiosyncrasy consistently inherent to the built form of cities within their countries and regions, captured by a probabilistic measure of entropy~\cite{brigattiEntropyHierarchicalClustering2021,nettoUrbanFormInformation2023} (Fig.~\ref{fig7} -- \textbf{cf. Chapter Z on architectural design and self-organization}).

\subsection{Shannon entropy as a measure of diversity} \label{city_diversity}
Entropy measures can be used as diversity indices, i.e. quantitative measures that may reveal how equiprobable specific features are in a distribution~\cite{brigattiEntropyHierarchicalClustering2021}. In general, spatial diversity measures focus on how a set of social variables, spatial characteristics, or urban attributes are evenly or unevenly distributed over landscapes and urban spaces. Applications in landscape ecology focus on patch mosaics related to size, density, fractal dimension, contagion, and nearest neighborhood distances~\cite{vrankenReviewUseEntropy2015}, taking into account the relative percentage of land-use types within areas~\cite{turnerLandscapeEcologyTheory2015,brownMixedLandUse2009}. In turn, entropy finds a broad range of applications as a diversity index in urban studies -- from street orientations and lengths seen in Sec.\ref{city_form} and urban parcel sizes~\cite{bitnerEntropyLandParcel2021} to the mixing of urban activities (land uses) or the segregation of social groups in urban spaces. We shall start this section by showing how Shannon entropy is applied to land-use mix (LUM).

\subsubsection{Entropy measures of land-use mix} \label{city_div_LUM}
\paragraph{}
LUM entropy measures application started from a transportation perspective in the late 1980s, mainly focused on the potential influence of land-use diversity over transport choices such as walking, cycling, and public transport~\cite{cerveroAmericaSuburbanCenters1988,frankImpactsMixedUsed1994}. LUM was extended into public health studies in the 2000s associated with walkability and health benefits~\cite{frankLinkingObjectivelyMeasured2005}. Despite the variety of diversity measures for land uses, entropy-based LUM indices are likely the most often-used measure of land-use diversity~\cite{ewingTravelBuiltEnvironment2010,songComparingMeasuresUrban2013a}. The concept of land-use mix implies that a distribution of a type of land use (e.g.\ retail or residential activities) across a limited spatial range is related to the distribution of other land uses. Cervero~\cite{cerveroAmericaSuburbanCenters1988,cerveroSuburbanEmploymentCenters1989} introduced Shannon entropy as a LUM measure, later employed by Frank and Pivo~\cite{frankLinkingObjectivelyMeasured2005} to analyze the impacts of local diversity over trips and transportation choices (cf.~\cite{kochelmanTravelBehaviorFunction1996,cerveroTravelDemand3Ds1997}). Frank et al.~\cite{frankLinkingObjectivelyMeasured2005} proposed a standardization of the measure dividing by a natural logarithm of the number of classes $J$ considered
\begin{equation}
H = -\sum_{i}\frac{P_{i}\ln(P_{i})}{\ln(I)}
\, 
\end{equation}
which reduces the influence of richness (the number of different land uses), emphasizing evenness (or unevenness) in the distribution of land use in an area. The standardization of Shannon entropy in land-use studies assumes that maximum entropy occurs when the percentages of each land-use distribution are exactly equal locally, reflecting a local best mix or balance of uses. Of course, the delimitation of areas of analysis according to administrative or analytic entities, such as radius-based buffers or street network distances, plays a relevant role in measurement. An apparently mixed local distribution of land uses can be dramatically different from distribution patterns over a whole city. 

Song et al.~\cite{songComparingMeasuresUrban2013a} propose an adaptation of diversity measures considering both local and global compositions, computing an entropy index relative to a `reference geography'. Local probabilities $P_{j}$ expressed in the Shannon equation do not refer to local percentages but the ratio of local and global percentages, generating a local quotient for each land use ($q$) with values greater than 1. Local distributions can be compared with the global distribution, computing a local diversity relative to the diversity measured at a global scale. In such an approach, the local entropy-based LUM measure intrinsically relates to diversity unevenness at the global scale -- something traditionally found in the dissimilarity index proposed by Duncan and Duncan~\cite{duncanMethodologicalAnalysisSegregation1955a} for estimating spatial segregation -- our next subject.

\subsection{Entropy and spatial segregation} \label{city_segreg}
Spatial segregation studies are based on the uneven locational distribution of homogeneous social groups (according to social characteristics, e.g.\ income, race, ethnicity and so on) across spatial areas, regions or places. Segregation is quite a diverse phenomenon that encompasses the separation of people in different situations of life, from institutions like schools to labor positions, but residential segregation is undoubtedly its most explored spatial domain. 
Similarly to land-use diversity, residential segregation was initially measured as the uneven locational distribution of population classified into social groups over a defined area. Since the 1950s, the favored measure was the dissimilarity index popularized by Duncan and Duncan~\cite{duncanMethodologicalAnalysisSegregation1955a}. A measure of entropy-based residential segregation only emerged two decades later, proposed by Theil and Finizza~\cite{theilStatisticalDecompositionAnalysis1972} -- capturing interest in the field~\cite{jamesMeasuresSegregation1985,whiteSegregationDiversityMeasures1986}. The entropy index $H$ measures evenness in residential location based on a weighted average deviation of the entropy in each spatially delimited area (e.g.\ census tract) from the entropy level of the city as a whole, therefore expressed as a fraction of the city's total entropy
\begin{equation}
H = -\sum_{i=1} \left(t_{i}\frac{\hat{H} - H_{i}}{\hat{H}}\right)
\, ,
\end{equation}
where $\hat{H}$ is the Shannon entropy of the distribution of social groups over the city as a whole, and $H_{i}$ is the local entropy for the same set of social groups. $H$ is related to the overall entropy of total area -- a measure of how more socially homogeneous (and segregated) residential areas are on average than the total population of the city. $H$ varies from 0 (residential areas with maximum social diversity or unevenness) to 1 (areas that have complete social homogeneity). 

\begin{figure}
    \centering
    \includegraphics[width=250pt]{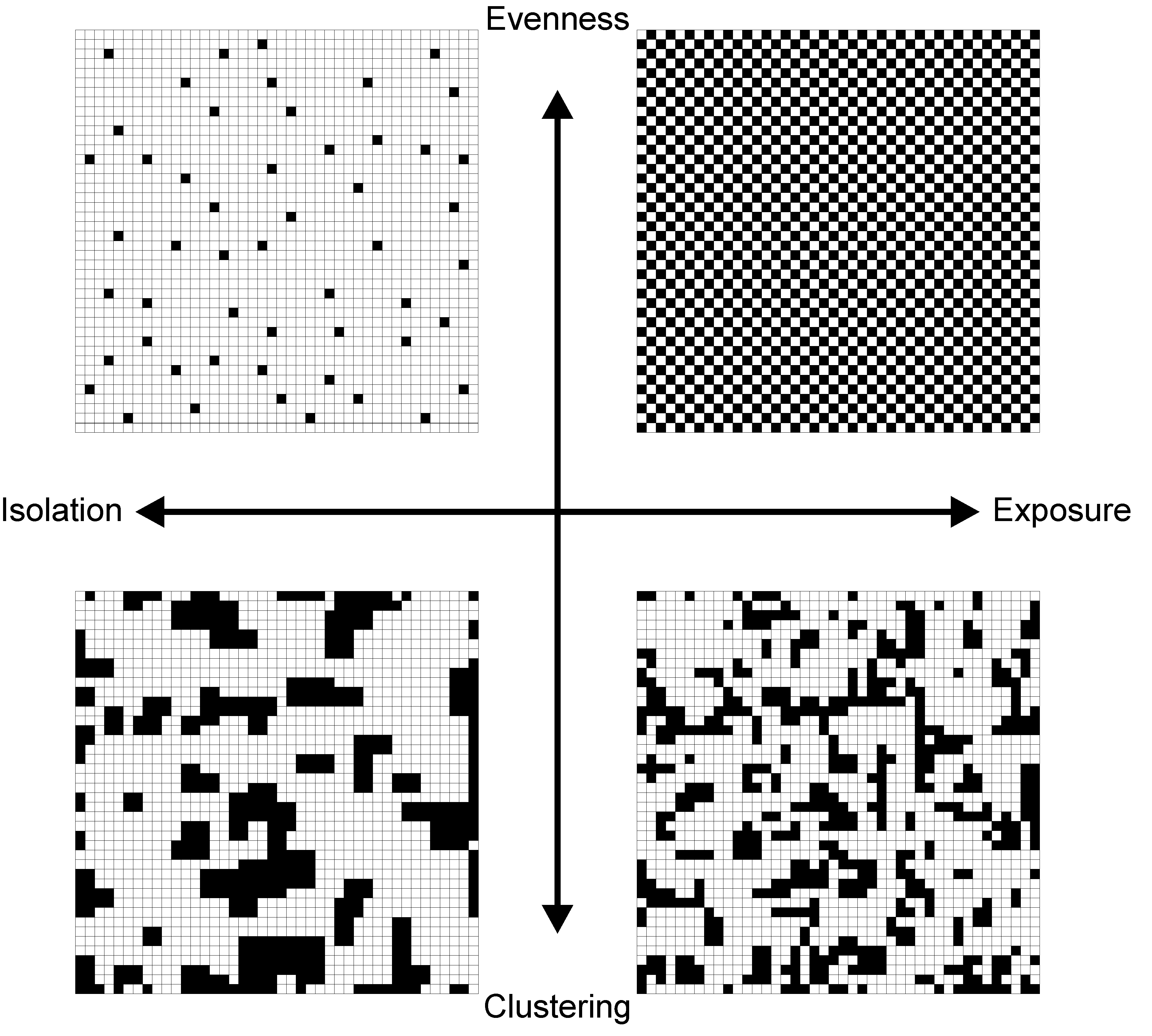}
    \caption{Distributions of black and white cells according to locational properties, from highly segregated to less or non-segregated distributions in a two-dimensional space (after \cite{reardonMeasuresSpatialSegregation2004a}).
    }
    \label{fig8}
\end{figure}

The entropy-based measure evolved since then through a set of adaptations as segregation measurement~\cite{whiteMeasurementSpatialSegregation1983a,openshawModifiableArealUnit1984} (\textbf{see Chapter Y}). Massey and Denton~\cite{masseyDimensionsResidentialSegregation1988b} identified five dimensions of measurement: evenness, exposure, concentration, centralization, and clustering. Reardon and O’Sullivan~\cite{reardonMeasuresSpatialSegregation2004a} readdressed the probabilistic isolation/exposure proposition by Bell~\cite{bellProbabilityModelMeasurement1954a} in 1954 to consider a simpler framework, i.e.\ spatial evenness and clustering (referring to the extent to which groups are segregated or not in urban space), spatial exposure (locations that favor encounters with members from another social group), and isolation (contact with members from one's own group) (Fig.~\ref{fig8}). 
Note that these measures oversimplify the possibilities of interaction between groups and do not effectively handle encounters or contact since such social variables are dependent on people's movement, which is absent from analyses of purely locational distributions~\cite{nettoSocialFabricCities2017}. Locational exposure should be understood as a potential for actual social exposure.

\subsection{Entropy as a means to understand urban size and densities, growth and sprawl} \label{city_sprawl}
\begin{figure}
    \centering
    \includegraphics[width=\textwidth]{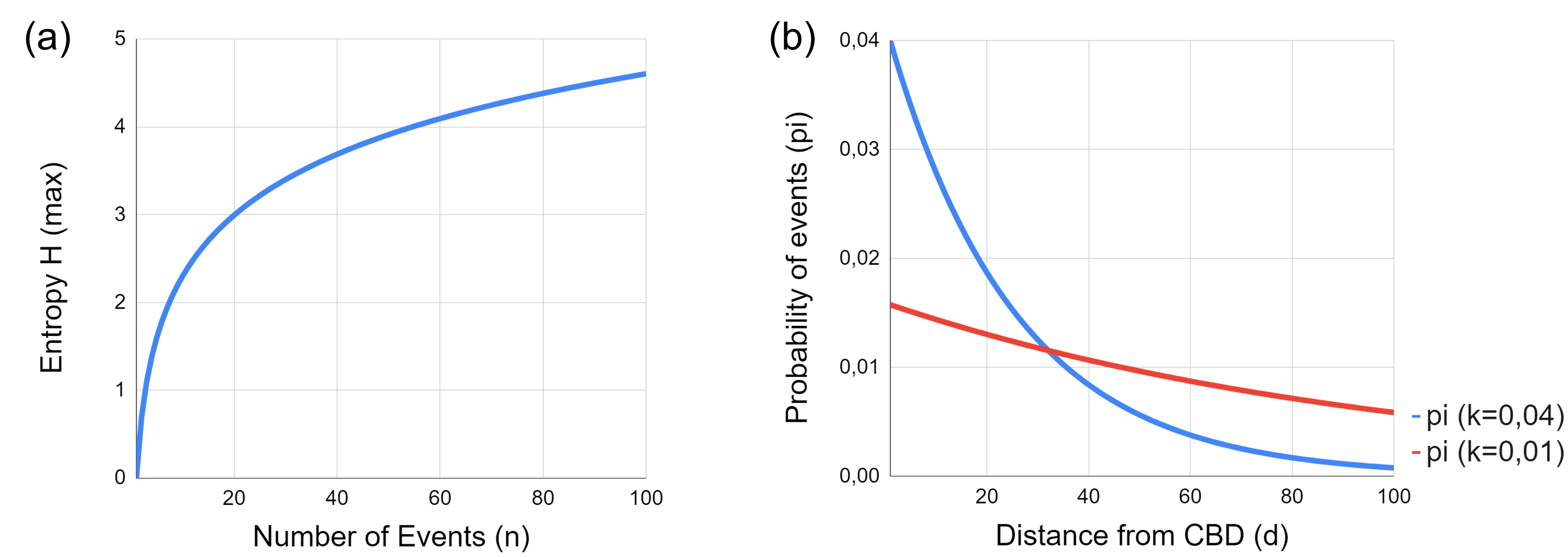}
    \caption{Entropy and urban densities varying with distance. (a) Entropy increases with the number of events; (b) probabilities decrease with distance in urban systems according to different energy costs to movement (after~\cite{battySpatialEntropyInformation2021}).
    }
    \label{fig9}
\end{figure}

A didactic way of seeing how entropy helps us understand urban densities is found in~\cite{battySpatialEntropyInformation2021}. Batty shows that entropy varies not only as distribution changes within a configuration but also as the number of objects or occurrences in the distribution changes, reflecting a trade-off between number and distribution (Fig.~\ref{fig9}a). The probabilities of discrete events occurring in a spatially extensional system like a city are not equally distributed across its space. Positions in geographical space are subject to differential potentials associated with their relative distance from other positions. The effect is a regular negative exponential distribution emanating from central positions -- to be sure, a distribution found in different urban phenomena such as densities of population and retail activities or the rent curves fundamental in urban economics. The exponential function can be steeper or flatter according to a friction-of-distance parameter $\beta$. A steeper curve indicates higher energy costs to the movement (Fig.~\ref{fig9}b -- \textbf{see Chapter M on mobility}). Such distributions have effects across virtually everything that happens in a city -- from the probabilities of encountering certain activities or people on the streets to the distribution of built form, economic diversity and intensities of exchange. As a result, a comparatively smaller number of locations concentrate probabilities of occurrence of such events. The blue curve in Fig.\ref{fig9}b reflects an internally differentiated distribution and contains more order than the flatter one (cf.\ Fig.~\ref{fig2}).

Urban densities and their different yet intrinsic levels of order may be clarified as probability distributions and associated entropy levels -- and they are inherently related to distance and therefore, city size. These properties are naturally useful to understand the implications of urban growth -- and different forms of growth, with distinct regularities in the density of land occupation, as suggested by the curves in Fig.\ref{fig9}b. An emblematic case of low-density growth is the so-called urban sprawl~\cite{ottensmannUrbanSprawlLand1977}. Urban sprawl has some key features, i.e.\ the amount of land that is built up, the dispersion of built-up areas over a landscape, and the density gradient of people living or working in these areas~\cite{nazarniaHowSuitableEntropy2019}. Methods for measuring urban sprawl cover a wide variety of aspects and landscape metrics, and have been used for its quantification~\cite{galsterWrestlingSprawlGround2001,tsaiQuantifyingUrbanForm2005}. Confronting a lack of indicators to measure urban sprawl~\cite{torrensMeasuringSprawl2000,yehMeasurementMonitoringUrban2001} proposed an effective measurement of urban sprawl using entropy focused on two main domains. 
(i) distance from urban central places and (ii) spatial patterns. Shannon entropy is used as a measure of spatial concentration or dispersion of a geographical variable $x$ among a number $n$ of zones. A series of concentric rings from the city's central place can be defined, and the density of urbanized areas in each buffer zone is computed. Urban concentrations tend to obtain lower values of relative entropy, and at the end of the spectrum, higher levels of sprawl are associated with higher values of relative entropy. 

Such measurement of sprawl as an entropic state of urban structures can account for time as the difference in entropy in $t$ and $t+1$ as an indicator of change, suggesting a city's trajectory towards either dispersion or compactness~\cite{yehMeasurementMonitoringUrban2001}. Shannon entropy has been widely used to understand sprawl dynamics in the last two decades~\cite{sudhiraUrbanSprawlMetrics2004,kumarSpatiotemporalAnalysisMonitoring2007,borrusoGeographicInformationAnalysis2013,aburasMonitoringAssessmentUrban2018a,nazarniaHowSuitableEntropy2019,dasAssessmentUrbanSprawl2021}, as a measure of the degree of concentration or dispersion of spatial variables sensitive to the variations in the shapes and sizes of urban areas~\cite{yehMeasurementMonitoringUrban2001}.

\subsection{Entropy and the self-organization of cities} \label{city_self-org}
The spatial production of cities through urbanization processes involves the emergence of structures. Similar to biological systems, cities are open, non-equilibrium systems functionally composed of different, specialized subsystems enacted by human and non-human agents, interacting and interchanging matter, information, and energy, cf.~\cite{rondinel-oviedoConstructionDemolitionWaste2023}. Cities as complex systems relate to entropy at different levels and moments of self-organization during their production and self-reproduction.
\begin{enumerate}
\item \textbf{The production of cities and urbanization processes.} A thermodynamic interpretation of the emergence of organized urban systems means that cities diminish their internal entropy while elevating the entropy of their surroundings. This will have implications for the survival and sustainability of cities, societies, and the environment itself (see Sec.\ref{city_sustainability}).
\item \textbf{The autopoiesis of cities.} A crucial question in the relationship of cities and entropy is how urbanization comes into being as a form of structuration countering tendencies towards entropy. Urban systems are characterized by specific and non-random structures. In the morphogenesis of structures emerging from possible configurations, from the micro to the macro structures, self-assembling urban systems restrict themselves to occupying a limited number of states. Through trial and error and other evolutionary processes (Sec.~\ref{city_form}), urban systems may take on configurations that minimize entropy production, cf.~\cite{prigogine1945moderation}. Cities are formed as specialized, mutually dependent subsystems like activity systems materialized in buildings and interconnected by street networks. In such autopoietic structuration~\cite{maturanaAutopoiesisCognitionRealization1991} across scales, subsystems become integrated into the functioning whole of the city.
\item \textbf{The workings of cities, or how urban systems survive.} Cities are systems that emerge as pockets of low entropy. Yet their own internal levels of entropy (endogenous entropy) can fluctuate while also exposed to random fluctuations in their external milieu (environmental or exogenous entropy). Entropy is produced inside and outside the system boundary~\cite{rondinel-oviedoConstructionDemolitionWaste2023} and relates to the loss of structure and organization. Cities require energy to synthesize and maintain their ordered structures. They respond to such losses and environmental fluctuations by adapting to changing conditions and improving their own structures.
\end{enumerate}

Cities resist a natural tendency to disorder in the form of degradation of their subsystems through agencies and mechanisms of self-repair, which is the subject of the following section.


\subsection{Entropy and the sustainability of cities and societies} \label{city_sustainability}
The connection between entropy and urban sustainability as the ability of cities to persist over time without threatening adjacent systems has been mostly explored in relation to the thermodynamical understanding of entropy. Such an ability evokes Schr\"odinger's~\cite{schrodingerWhatLifePhysical1944} notion of delayed entropy, a property of systems capable of temporarily delaying the natural process leading them toward maximum entropy. This delay is achieved through interactions with the environment, extracting negative entropy or free energy~\cite{fristonFreeEnergyPrinciple2012}. The interaction of complex urban systems with their environment, while essential for survival, is a `double-edged sword'. It is both the source of the system's survival and the force responsible for its eventual decay into maximum entropy. Once a system can no longer extract negative entropy, it succumbs to disintegration by the environment~\cite{portugaliSchrodingerWhatLife2023}.  In short, like other forms of human settlement, e.g. rural or non-urban spaces, the city needs a constant supply of energy to keep its entropy low. However, the amount of energy required in ever-growing economic systems materialized in the form of cities runs faster against the regenerative processes of the environment and its
 natural limits as an energy resource. 
\begin{itemize}
\item[$\bullet$] \textbf{Energy use and entropic outputs during urban metabolism.} An open self-organized system needs to import available energy and material from their host environments in order to reproduce itself~\cite{rondinel-oviedoConstructionDemolitionWaste2023}. Processes of urbanization, such as urban growth and sprawl, and the daily workings of cities involve intensive use and transformation of resources, materials and energy, leading to negative externalities such as waste, effluents, and pollution. Such transformations are analogously found in biological systems (Sec.\ref{hist_biol}).
\item[$\bullet$] \textbf{Cities as contributors to increasing environmental entropy.} Cities are recognized as significant contributors to the environmental crisis. Their exogenous consumption and externalities are directly linked with entropic environments. In urban growth and sprawl (Sec.\ref{city_sprawl}), entropy must be kept within a range defined by a minimum value below which the system becomes vulnerable and a maximum value above which the system becomes unsustainable~\cite{cabralEntropyUrbanSystems2013}.
\end{itemize}

Entropy appears frequently described in the urban science literature as an opposite force to sustainability~\cite{rondinel-oviedoConstructionDemolitionWaste2023}. However, a direct opposition between entropy and sustainability is an oversimplification. Societies and cities also increase their internal entropies as an effect of the daily creation of new interactions, social relationships, groups and agencies, and through eventful cognitive, social, and technological innovations. Entropy increases as our everyday actions and interactions in the city go into momentary states of unpredictability and new possibilities are presented to us~\cite{luhmannSocialSystems1995,nettoSocialInteractionCity2017}. These serendipitous processes may trigger positive entropic effects -- even though they may be accompanied by initially negative ones such as the disruption of existing agencies, like the downfall of firms or whole industries. Furthermore, cities have become a crucial means to a broader sense of societal sustainability. Societies reproduce themselves subject to limited resources, fluctuating environments, external shocks and systemic crises, all contributing to uncertainty and hindering decision-making. The nature of these challenges can be clarified through the concept of entropy. Societies face entropy all the time (Sec.\ref{hist_social}). While preparing for the negative effects of entropy is vital to how societies become resilient, fostering positive entropic effects triggered by innovation and serendipity is crucial for creative societies. Cities are key for both processes. Societies endure and manage different forms of entropy through three continuous processes.
\begin{enumerate}
\item \textbf{Creating physical and semantic urban structures.} To support volatile interaction systems, societies shape their built environment locally via autopoietic processes~\cite{maturanaAutopoiesisCognitionRealization1991}, self-organizing as they arrange physical space in the form of cities through morphogenesis and the semantization of the urban form, i.e. the creation of social contents, meaning and information associated with physical objects such as buildings and places~\cite{nettoSocialFabricCities2017}.
\item \textbf{Exploring urban structures} as they encode information critical to people’s lives, daily interactions and collective cooperation. People living in cities deal with daily activities and challenges through cognitive and practical mechanisms like retrieving and enacting information from their built environment~\cite{nettoCitiesInformationInteraction2018,haken2015information}.
\item \textbf{Stressing and improving urban structures} in response to systemic fluctuations through properties like anti-fragility~\cite{talebAntifragileThingsThat2012} and self-improvement mechanisms, like bottom-up collective learning and top-down monitoring capacities in planning. Such capacities trigger and guide recursive changes in the built environment, increasing urban resilience and keeping social entropy at bay through continuous works of adaptation in cities.
\end{enumerate}

Cities can improve or hinder social dynamics, reducing or expanding societal demands and externalities because cities have structures that enable and shape such dynamics. Theoretically, two cities with the same population number and geographical size would perform differently if their internal structures were different -- say, a fragmented urban structure that makes navigation difficult, disturbing interactions and triggering diseconomies, or a structure containing internal correlations capable of offering spatial legibility and fluency to daily movement and interaction. Urban structures, therefore, matter. They support the challenging passage from individual actions to large-scale interaction systems, a crucial recursive step in how societies manage entropy~\cite{Netto_Meirelles_Ribeiro_2020}. Cities are vital in that sense.


\section{Criticism} \label{criticism_4}
The concept of entropy is often misunderstood both in general and within the specific context of urban science -- for some possible reasons.
\begin{itemize}
\item[$\bullet$] \textbf{Complexity and abstraction.} Entropy is a highly abstract and counter-intuitive concept rooted in statistical mechanics and information theory. Its application involves mathematical formulations and probabilistic reasoning, making it challenging for researchers without a background in these fields to grasp its nuances. This is particularly the case in urban studies, which often involve researchers with backgrounds in fields such as sociology, geography, and planning. Without interdisciplinary training that includes concepts from physics and information theory, researchers may struggle to appreciate the full scope of the concept and its applications.
\item[$\bullet$] \textbf{Urban dynamics are multifaceted.} Applying entropy in urban science involves deals with the multifaceted dynamics of cities, which are influenced by social, economic, and environmental factors. The complexity of urban systems can make it challenging to isolate and quantify the specific contributions of entropy, leading to potential misinterpretations.
\item[$\bullet$] \textbf{Inconsistencies in the evolution and interpretations of the concepts over time.} The concept of entropy has evolved over time since its roots in classical thermodynamics. This evolution can lead to varying interpretations, especially when researchers are exposed to different formulations of the concept. The diversity of applications in urban studies can contribute to confusion, as researchers may find entropy in different forms and contexts \cite{battySpatialEntropy1974a}. For instance, an association with disorder in the everyday sense can lead to oversimplification.
\end{itemize}

While the concept of entropy has been applied in urban studies and city science, there are critiques associated with its use and the application of associated measures.
\begin{itemize}
\item[$\bullet$] \textbf{Assumption of equilibrium.} Entropy models often assume equilibrium or quasi-equilibrium conditions~\cite{battySpatialEntropy1974a,wilson2011entropy}. However, cities are dynamic and rarely in a state of equilibrium. This assumption may not fully capture the ongoing evolution and adaptation of urban systems~\cite{batty2012origins}.
\item[$\bullet$] \textbf{Single-domain approaches.} Some entropy measures focus on a single domain, such as land use or population distribution. This narrow focus may overlook the complexity of urban systems. 
\item[$\bullet$] \textbf{Overemphasis on physical structures.} Some entropy-based approaches emphasize the physical form of cities at the expense of its relations to semantic aspects of urban form.
\item[$\bullet$] \textbf{Limited relations to social processes.} Entropy measures are frequently applied to spatial distributions. 
Connections to volatile dynamics of social interactions as intrinsic to the urban phenomenon and its complexity are frequently overlooked, along with the effects of top-down social agencies. Entropy-maximizing input-output frameworks \cite{wilsonEntropyUrbanRegional1970,wilson2011entropy} are exceptions but do not encompass the richness of social interaction.
\item[$\bullet$] \textbf{Overlooked temporal processes.} Many entropy-based approaches do not adequately consider temporal dynamics. Entropy measures applied in analyses of urban systems states may easily ignore changes and adaptations occurring within cities \cite{leibovici2014local}.
\item[$\bullet$] \textbf{Exclusion of contextual information.} Context-specific factors, such as local policies, cultural nuances, and historical legacies, may not be adequately integrated into entropy calculations. Investigating the generalizability of entropy measures in various urban settings is an ongoing challenge. 
\item[$\bullet$] \textbf{Lack of standardization.} There is a lack of standardized methods for calculating entropy in urban studies. Different researchers may use different measures, making it challenging to compare findings and hindering the establishment of consistent benchmarks \cite{reggianiHandbookEntropyComplexity2021}.
\item[$\bullet$] \textbf{Data resolution issues.} The resolution of data used to calculate entropy can impact the results. Coarse-grained data may overlook fine-scale variations, while fine-grained data might introduce noise and overemphasize local patterns \cite{brigattiEntropyHierarchicalClustering2021}. Choosing an appropriate level of resolution is a crucial but challenging task.
\item[$\bullet$] \textbf{Data acquisition challenges.} Obtaining accurate and comprehensive urban data for entropy calculations can be challenging. Data gaps, inaccuracies, and the dynamic nature of cities also pose significant obstacles.
\item[$\bullet$] \textbf{Limited predictive power.} Entropy models may have limited predictive power. Cities are influenced by unpredictable factors, and deterministic approaches might not account for the emergent properties of cities in response to such factors \cite{batty2012origins}.
\end{itemize}

\section{Open questions} \label{open_quest_5}
The use of the concept and measures of entropy in urban studies and city science has advanced our understanding of complex urban systems, but several open questions and challenges remain. Some of the key issues and problems yet to be fully addressed and explained include:
\begin{itemize}
\item[$\bullet$] \textbf{Multi-domain entropy.} The majority of entropy measures in urban studies focus on a single domain, such as land use or population distribution. Developing comprehensive multi-domain entropy measures that capture the complexity of urban systems in terms of social, economic, and environmental factors remains a challenge \cite{reggianiHandbookEntropyComplexity2021}.
\item[$\bullet$] \textbf{Dynamic interaction networks.} Cities operate as dynamic interaction networks \cite{nettoSocialFabricCities2017}, and current entropy measures may not fully capture the evolving relationships between different elements in the urban environment. Developing entropy measures that consider the multi-layered and hybrid network structures inherent in urban systems is a key step in research \cite{battySpatialEntropyInformation2021}.
\item[$\bullet$] \textbf{Multi-scale framework.} Entropy values can be scale-dependent, and there is a need to develop frameworks for interpreting entropy at various spatial and temporal scales. Understanding how entropy values change based on the chosen scale is an area for further exploration \cite{battySpaceScaleScaling2010,leibovici2014local}.
\item[$\bullet$] \textbf{Integration of qualitative or semantic information.} Quantifying information in the context of cities is challenging. Unlike physical systems, information in urban settings is often qualitative, such as cultural richness, semantic information in space retrieved and enacted by agents~\cite{nettoSocialInteractionCity2017,nettoCitiesInformationInteraction2018} and converting it into quantifiable measures for entropy calculations can be problematic. Promising insights may be gained from exploring methods to integrate qualitative and semantic information into entropy measures~\cite{haken2015information} and acknowledging the subjective aspects of the urban experience.
\item[$\bullet$] \textbf{Policy and planning implications.} Bridging the gap between entropy measures and practical policy and planning decisions is an ongoing challenge. 
The mechanisms and thresholds that contribute to or hinder urban resilience~\cite{cabralEntropyUrbanSystems2013}, how entropy influences a city's ability to adapt and improve from environmental and internal fluctuations, and how entropic forces and their effects on different communities are fresh lines of inquiry. Exploring entropy measures to assess spatial scenarios as performance indicators to inform sustainable urban development and governance remains an important avenue for future research~\cite{rondinel-oviedoConstructionDemolitionWaste2023,10.1007/s10661-007-0089-1,10.1007/s00477-011-0493-5}.
\end{itemize}

\section{Summary} \label{summary_6}
In this chapter, we followed the historical trajectory of the concept of entropy from its inception in thermodynamics and statistical mechanics in the last decades of the nineteenth century to some of its latest incarnations to understand cities as complex systems. We attempted to demystify Boltzmann’s understanding of the probability of different molecular arrangements coarse-graining into macrostates and Gibb’s function of the probability distribution over phase space.
We visited Shannon’s definition of entropy in information theory as a measure of uncertainty in the engineering of communication channels and the transmission of data in the mid-twentieth century.
Evoking Prigogine~\cite{prigogine1989entropy} again, those pioneering visions of a very strange yet fundamental concept to understand the world transcended their original fields into a rich multidisciplinary tapestry of scientific applications. 

Explorations in urban science were certainly enriched by how the concept was interpreted in biology -- from the conditions for the origin of life as organized living systems and their metabolism to their recursive adaptation in the face of changing environments. 
The subsequent absorption of the concept in the social sciences threw light on how societies manage internal and external entropies. From the point of view of entropy and complexity, a key process is how microscopic actions are assembled into possible functional configurations. That is the passage from possibilities and uncertainties surrounding people's interactions to large-scale economic and political organization. Institutionalized structures like customs, rules, and culture have a role in generating such organized states~\cite{parsonsSystemModernSocieties1971,giddens2004constitution}. 

So have cities. Somewhere between structure and randomness, cities are more than an environment to the always ongoing emergence of social structures. Societies and cities are rather \emph{co-evolving adjacent systems}. 
The city is what Parsons~\cite{parsonsSystemModernSocieties1971} called an `action frame of reference' -- a connective fabric that informs, constrains and enables interaction choices~\cite{nettoSocialInteractionCity2017}.
Thus, we explored the concept of entropy to understand cities, from the conditions of their emergence and evolution to the challenges in their continuity.
We searched for approaches, measures and empirical applications of the concept to understand cities as entropy-maximizing interaction systems, along with urban form, the distribution and diversity of land uses, and patterns of spatial segregation and densities, growth and sprawl. 

We looked into the self-organization of cities through the lens of entropy. From the space of possible configurations, the emergence of urban patterns might also minimize the rate of entropy in the workings of cities themselves, including the assembling of large-scale interaction systems out of individual actions ~\cite{nettoCitiesInformationInteraction2018}. 
Finally, we described entropy as an aspect of sustainability, seeing the role of cities in how societies navigate entropy by creating urban structures, exploring such structures in daily interactions, and stressing and improving them in response to internal and external fluctuations.

The concept and measures of entropy provide valuable insights into critical aspects of urban systems and urban life. Exploring such richness and addressing the challenges to integrate entropy approaches with broader perspectives considering the diverse and dynamic nature of cities outlined above seem essential steps for a more comprehensive understanding of urban complexity.

\bibliographystyle{splncs04}
\bibliography{Refs_Entropy_and_Cities_2}

\end{document}